\documentclass{aa}  

\usepackage{txfonts}
\usepackage{graphicx}	
\usepackage{graphicx}	
\usepackage{amssymb}	

\usepackage{epstopdf}
\usepackage[font=footnotesize,labelfont=bf]{caption}
\usepackage{color}
\usepackage{amsmath}
\usepackage{mathtools}
\usepackage{booktabs,caption}
\usepackage[flushleft]{threeparttable}
\usepackage{hyperref}
\usepackage{csquotes}
\defcitealias{TM99}{TM99}
\usepackage{enumitem} 
\setlist[itemize]{label=$\bullet$} 

\usepackage{perpage}
\def\apjs{{\rm ApJS}}                  
\def\apjs{{\rm ApJ}}                   
\def\apjl{{\rm ApJ Let}}               

\def\a4{\hsize 17.0cm \vsize 25.cm}

\begin{document} 
   \title{The bright, dusty aftermath of giant eruptions \& H--rich supernovae}
   \subtitle{Late interaction of supernova shocks \& dusty circumstellar shells 
   }
   \titlerunning{The bright, dusty aftermath of giant eruptions \& H--rich supernovae}

\author
{Diana B. Serrano-Hern\'andez$^{1},$
Sergio Mart\'{\i}nez-Gonz\'alez \thanks{Corresponding author. E-mail: sergiomtz@inaoep.mx}$^{1},$ 
Santiago Jim\'enez$^{2},$
Sergiy Silich$^{1},$
Richard W\" unsch$^{2}$}

   \institute{Instituto Nacional de Astrof\'\i sica \'Optica y Electr\'onica, AP 51, 72000 Puebla, M\'exico
         \and
             Astronomical Institute of the Czech Academy of Sciences,  Bo\v{c}n\'\i\ II 1401/1, 141 00 Praha 4, Czech Republic}

   \date{Received: ; Accepted:}
\authorrunning{Serrano-Hern\'andez et al.}
 
\abstract
  {The late-stage evolution of massive stars is marked by periods of intense instability as they transit towards their final core-collapse. Within these periods, stellar eruptions stand out due to their hallmark of exceptionally high mass loss rates, resulting in the formation of copious amounts of dust. However, the survival of these dust grains is threatened by the powerful shock waves generated when the progenitor star explodes as a supernova (SN).} 
  {We aim to assess the impact of  selected cases of hydrogen--rich SN explosions from progenitors of 45, 50, and 60 M$_\odot$ on dust grains formed after giant stellar eruptions, exploring late interactions with circumstellar shells that occur a few years to centuries after the eruption.}
  {We present 3D hydrodynamical simulations that follow the evolution of dust particles in a scenario that includes, for the first time, the progenitor's stellar wind, a giant stellar eruption, and the eventual SN explosion, while in line with the mass budget prescribed by stellar evolutionary models.}
    {For a standard SN ejecta mass of $10 \ \textrm{M}_\odot$, kinetic energy of $10^{51}$ erg, and a long 200-year eruption-SN gap, only 25\% of the dust mass remains 250 years post-explosion in a spherical CSM, and only 2\% a century after the explosion in a bipolar CSM. Conversely, a shorter gap of a dozen years preserves 75\% of the dust mass after shock-processing for a standard explosion, while this drops to 20\% for more massive ($15$–$20 \ \textrm{M}_\odot$) ejecta with kinetic energy of $5 \times 10^{51}$ erg.}
   {The CSM geometry and an early SN remnant transition to a radiative phase impact dust survival. As the shock wave weakens from efficiently converting kinetic energy into thermal radiation (up to half of the injected kinetic energy), there is a greater potential of survival, not only for dust in the CSM but also for SN-condensed dust (due to a weaker SN reverse shock), and pre-existing dust in the ambient ISM. Against expectations, a larger fraction of the dust mass can survive if the SN occurs just a few years after the eruption event.}


   \maketitle
\section{Introduction}
Interstellar dust plays a crucial role in the interstellar medium (ISM), influencing various astrophysical processes that impact the gas dynamics, chemistry, and thermodynamics within galaxies. These processes include the formation of molecules, planets, and stars \citep{Draine03-a, Tielens2005, Draine2011}. The dust ability to absorb and scatter starlight is most pronounced in UV and optical regimes, while it emits radiation in the infrared regime \citep{Draine03-b}.
\begin{figure*}
    \centering
    \includegraphics[width=0.8\textwidth]{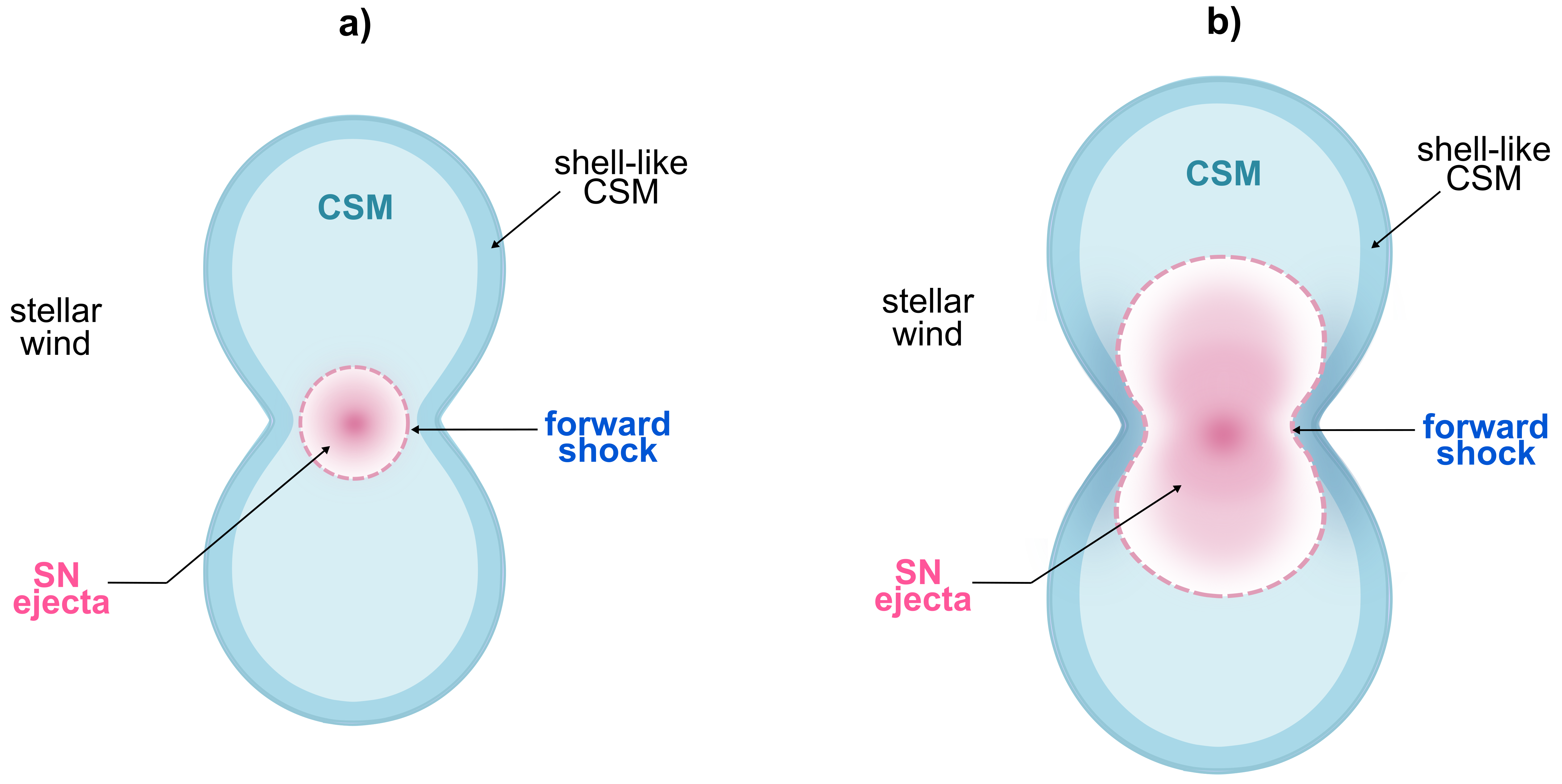}
    \caption{Schematic (not to scale) of a supernova (SN) remnant evolving within a bipolar circumstellar medium (CSM) created by a stellar eruption, subsequent to the ejection of the stellar wind. Panel a) shows the expansion of the SN ejecta and the forward shock at early times after the explosion. Panel b) presents the time when the SN begins to interact with the densest medium left by the eruption. }
    \label{csm-sn}
\end{figure*}
The origin and nature of dust in galaxies remain not fully understood, although it is recognized to form in the dense and cold envelopes of evolved stars through their stellar winds, especially stars in the Asymptotic Giant Branch (AGB) stage, Red Supergiants, Wolf Rayet and Luminous Blue Variable (LBV) stars \citep[e.g.][]{Todini&Ferrara2001, Crowther2003, Masseyetal2005, Tielens2005, Dwek05, Cherchneff2013, Smith2010, Kochanek2011}. Furthermore, the ejecta of core-collapse supernovae are also known to have the required conditions to form dust and contribute significantly to the dust budget in the Universe \citep[e.g.][]{Elmegreen1981, Todini&Ferrara2001, Dunetal2009, Matsuurasetal2009, GallandHjorth2018, Sarangi&Matsuura18, Kirchschlageretal2019,Kirchschlageretal2020, Priestleyetal2021, Priestleyetal2022}.

During their evolution, stars experience fast outflows which inject mechanical energy, mass and momentum into the ISM. These outflows, known as stellar winds, lead to the formation of bubbles around the progenitor stars since the winds are ejected into the ISM at supersonic speeds \citep{Weaveretal1977,Dyson1989,vanMarleetal2004,Draine2011}.
Moreover, massive stars can also lose mass during eruptive processes, which tend to occur when the star becomes unstable as it approaches the so-called Eddington limit \citep[e.g.][]{Humphreys&Davidson1994}. Indeed, when the Eddington limit is reached, the radiation forces at the surface prevail over gravity, resulting in a substantially increased mass loss rate. 
These outbursts or stellar eruptions are explosive events during which a huge amount of gas is ejected from the star's surface into the surrounding medium. Therefore, they can lead to the creation of intricate shell-like structures and contribute to the enrichment of the ISM with heavy elements that can subsequently condense onto dust particles \citep{Kochanek2011, Smith14}. These eruptions can endure for months or even years, leading to the development of a dense circumstellar medium (CSM) surrounding the star \citep[e.g.][]{Smith2011, Davidson2020}. Previous numerical simulations have successfully modelled the formation and evolution of such CSM structures around massive stars \citep[e.g.][]{Garcia-Seguraetal1996b,Garcia-Seguraetal1996a}.

Furthermore, the powerful shock waves generated by core-collapse supernovae (SNe) can dramatically influence the fate of the dust grains formed during the preceding stellar outbursts, as they interact with pre-existing CSM weeks to centuries post-explosion \citep{Smith&Andrews2020}. The extreme temperatures, pressures, and velocities associated with the shock waves, along with the intense shock breakout radiation, can result in the destruction, fragmentation, or vaporization of the dust grains \citep[e.g.][]{Nozawaetal2010,Temimetal2015,GallandHjorth2018,Micelottaetal2018}.

These interactions often involve an asymmetric environment around the progenitor, typical of evolved massive stars, with non-spherical CSM structures like disks or bipolar outflows \citep{Smithetal2018,Qinetal2023}. This asymmetry strongly impacts the observed SNR-CSM interaction signatures and can result in a large amount of radiated energy when the SN blast wave interacts with the dense stellar wind and outburst-produced CSM \citep{vanMarleetal2010,Panetal2013,Vlasisetal2016,Smith2017, Smith&Andrews2020,Margalit2022, Martinezetal2023, Berstenetal2024}. 

The best known case of a stellar eruption in the Milky Way is the Great Eruption of $\eta$ Carinae ($\eta$ Car) around 1843, that formed its remarkable bipolar nebula known as the Homunculus \citep{Humphreys&Davidson1994, MacLowetal1996, Davidson&Humphreys1997}. $\eta$ Car is a luminous binary star system classified as LBV, with two massive components of about 90 M$_\odot$ and 30 M$_\odot$, respectively \citep{Maduraetal2012}. The Great Eruption resulted in the ejection of a massive amount of material, hence the high density, low temperature, and abundance of heavy elements in the erupted matter provided favorable conditions for the formation of dust particles \citep{Humphreys&Martin2012, Weis2012}. Indeed, up to 0.4 M$_\odot$ of dust were produced after the eruption \citep{Gomezetal2010}, raising questions about its fate in the event of a SN explosion \citep[e.g.][]{GallandHjorth2018}.     

Furthermore, dust formation in SNe showing interaction with a dense CSM has also been recently studied, providing both theoretical and observational evidence of the formation of new dust grains following the collision of the SN blast wave with the dense CSM \citep{SarangiandSlavin2022, Smithetal2023}.

Examples of core-collapse SNe exhibiting evidence of strong interaction with a dense CSM include SN 2005gj \citep{Trundleetal2008}, SN 2005gl \citep{GalYametal2007}, SN 2017hcc \citep{Kumaretal2019, Smith&Andrews2020}, SN 2005ip \citep{Smithetal2017}, SN 2015da \citep{Smithetal2023}, SN 2007od \citep{Andrewsetal2010},  SN 2004et \citep{Szalaietal2021}, SN 2013ej  \citep{Mauerhanetal2017, Szalaietal2021}, and SN 2018lab \citep{Pearsonetal2023}.

The wide range of timescales, from weeks to millennia, between eruptive episodes and subsequent supernova explosions, as highlighted by available data \citep{Brethaueretal2022}, is an intriguing research subject. This work examines these timescales as crucial parameters influencing the dynamics of these phenomena.

\vspace{2mm}
Here, we employ detailed 3D hydrodynamical calculations to investigate the impact of selected cases of hydrogen--rich SN explosions on dust grains formed during late-stage stellar outbursts. Our goal is to determine whether these explosions result in the complete or partial destruction of the circumstellar dust. In particular, we simulate the evolution of the SN blast wave, its interaction with the pre-SN erupted matter and stellar wind, and calculate on-the-fly the resulting destruction of the erupted dust grains. By comparing the initial and final amount of the dust grains, we quantify the degree of destruction and gain insights into the survival mechanisms of interstellar dust. 

\vspace{2mm}
The paper is organized as follows. Section \ref{section2} describes the implemented hydrodynamical scheme, and the initial conditions for the stellar winds (subsection \ref{sec_wind}), the erupted CSM evolution (subsection \ref{sec_er}), and the supernova remnant (SNR) evolution (subsection \ref{sec_sn}). In Section \ref{results} we present the results of the simulations, and the summary and conclusions are presented in Section \ref{conclusions}.

\section{Numerical setup} \label{section2}
Fig. \ref{csm-sn} presents a schematic illustration of the late-stage evolution of a massive star undergoing several mass-loss episodes. First, the fast stellar wind from the star continuously injects mass and energy into the ambient ISM gas. Later, the star goes through an eruptive episode that creates a bipolar CSM (shown in the scheme as blue). Finally, the star explodes as SN, expanding into the CSM (panel $a$). The SN forward shock eventually collides with the densest sections of the erupted matter (panel $b$), first towards the equator and later with the poles of the structure. 

\vspace{2mm}

Stellar winds may be inhomogeneous, which leads to a reduction in the empirical mass-loss rates \citep[][]{Oskinovaetal2007}. Therefore, we estimated the characteristic cooling timescale in the free wind regions (see Appendix \ref{Appendix0}, equation \ref{eq:tcrit}) and found that in all our models, it is much longer than our simulation timescales. However, it depends on the injection radius $R_w$, which we cannot decrease significantly. Another important issue is the potential impact of numerical diffusion, which can suppress or limit the growth of thermal instabilities.

Therefore, we have carried out a set of 3D hydrodynamical simulations using the adaptive-mesh refinement (AMR) code FLASH 4.6 \citep{Fryxelletal2000}. The code implements the Piece-wise Parabolic Method (PPM) to solve the hydrodynamic equations \citep{ColellaandWoodward1984}. The package PARAMESH \citep{MacNeiceetal1999} is used for implementing the AMR grid, which employs a block-structured adaptive mesh refinement scheme.

The simulation scheme includes the calculation of a time-dependent stellar wind (section \ref{sec_wind}) and the SN explosion (section \ref{sec_sn}) using the Wind module \citep[]{Wunschetal2008, Wunschetal2017}, which was also adapted to allow the injection of the stellar eruption (section \ref{sec_er}). Furthermore, we also include radiative cooling at solar metallicity as prescribed by \cite{Schureetal2009} assuming collisional ionization equilibrium and instantaneous electron-ion energy equipartition. 
In our case of study, the validity for these assumptions stems from the high densities and short timescales characterizing the region behind the forward shock, where the bulk of radiative cooling occurs \citep{Spitzer1962,SmithandHughes2010,Wongetal2016}. 
The implementation also includes the injection and destruction of dust particles, and the additional radiative cooling resulting from gas-grain collisions, calculated on-the-fly with the \textsc{Cinder} module \citep[Cooling INduced by Dust \& Erosion Rates,][]{MartinezGonzalezetal2018,MartinezGonzalezetal2019,MartinezGonzalezetal2022}. 

In our calculations, we assume three progenitor stars with masses of 45, 50 and 60 M$_{\odot}$ at the Zero Age Main Sequence (ZAMS). The first two progenitors are assumed to have formed with a metalliciy of 0.02 Z$_{\odot}$, whereas the third is assumed to have formed with the Milky Way's metalliciy of 0.73 Z$_{\odot}$. According to the Bonn Optimized Stellar Tracks \citep[BoOST][]{Szecsietal2022}, such stars evolve over to the point of forming a carbon core. This phase is reached when each star is about 99\% through their lifetime, having shed a substantial amount of mass through a stellar wind.
Despite this loss, the 45-M$_\odot$, 50-M$_\odot$ and 60-M$_\odot$ stars still maintain a mass $\sim 42.5$, $47$, and $37$ M$_\odot$, respectively.
In the remaining 1\% of the stars’ lifetime, we assume the occurrence of a giant stellar eruption, shedding $\sim25$ M$_\odot$ of material \citep[e.g.][]{Woosleyetal2007,Smithetal2010,Wangetal2022}. Subsequently, as the stars progress towards their ultimate fate, a hydrogen--rich core-collapse SN occurs \citep[consistent with expectations for carbon core masses below 40 M$_\odot$][]{Szecsietal2022}, ejecting $\geq 10$ M$_\odot$ of material and leaving behind a compact stellar remnant. Hydrogen--rich core-collapse SNe typically yield relatively small amounts of \(^{56}\textrm{Ni}\) \citep{Anderson2019}. Nevertheless, for energetic hydrogen--rich SNe, a higher $^{56}$Ni content is expected, as exemplified by the SNe iPTF14hls \citep{Wangetal2022}, and 2006gy \citep{Moriyaetal2013}. We thus  also account for the impact of radiative heating from the decay of \(^{56}\textrm{Ni}\) in the SN remnant evolution.

\subsection{Stellar Winds}\label{sec_wind}

An isotropic stellar wind is simulated according to the implementation of \cite{Wunschetal2017}, in which the wind mass and energy are inserted within a radius $R_{\textrm{w}}$, centered at the origin of the spatial domain. For each grid cell within $R_{\textrm{w}}$, mass is first added and the velocity is corrected to ensure momentum conservation. Subsequently, energy is also added according to the wind energy deposition rate. The gas density around the source, within $R_{\textrm{w}}$, is set as:

\begin{equation}\label{wind_who}
\rho_{w}(r) = \frac{\dot{M}_{w}}{4\pi v(r)r^2}, 
\end{equation}
and
\begin{equation}
v(r) = v_{\infty}r/R_{\textrm{w}},  
\end{equation}
respectively. In these equations, $\dot{M}_{w}$ is the time-dependent mass loss rate, $v_{\infty}$ is the wind terminal velocity and $r$ is the distance from the center. The wind insertion radius $R_{\textrm{w}}$ is set as small as possible such that the wind is approximately spherical.

The wind temperature, $T_{w}$, is set to $10^4$ K as in \cite{MartinezGonzalezetal2019}. Following \cite{Franketal1995,Gonzalez2022}, our calculations assume $\dot{M}_{w} = 10^{-3}$ M$_\odot$ yr$^{-1}$, and $v_{\infty}= 250$ km s$^{-1}$ around the time the subsequent eruption occurs. No dust is inserted together with the stellar wind as our aim here is to study the fate of the dust injected by the stellar eruption. The calculations assume solar metallicity ($Z=Z_{\odot}$).
 
The stellar wind is let to evolve until it extends beyond the boundaries of the computational domain. Afterwards, the stellar wind is switched off upon the insertion of the stellar eruption.

It is worth noting that the computational cost of the stable stellar wind phase was relatively low compared to the more demanding calculations for the following stages, which include both the stellar eruption and the SN remnant. This allowed us to include the wind phase in the simulations without significantly increasing the overall computational expense. We would also like to highlight that calculating the wind in 3D directly on the Cartesian grid avoids the potential sources of noise and inaccuracies that could arise from remapping a spherically symmetric wind solution onto a Cartesian grid.

\subsection{Erupted CSM evolution} \label{sec_er}

Here, the evolution of the expanding circumstellar matter resulting from a stellar eruption is described by considering several input parameters, including the deposited kinetic energy $E_{\textrm{k,e}}$, the maximum expansion velocity $v_{\textrm{max,e}}$, and the total mass of the eruption ejecta $M_{\textrm{e}}$. Similarly to the stellar wind insertion, the eruption ejecta is modelled by inserting mass and internal energy in all grid cells within an insertion radius during a single time-step. The radius of the inserted eruption, denoted as $R_{\textrm{e}}$, is limited by the spatial resolution. We minimize the insertion radius to cover the smallest possible volume so the eruption maintains an approximately spherical shape and ensures minimal error in the mass and energy insertion.
\begin{figure}
    \centering
    \captionsetup{skip=1pt}    \includegraphics[width=0.48\textwidth]{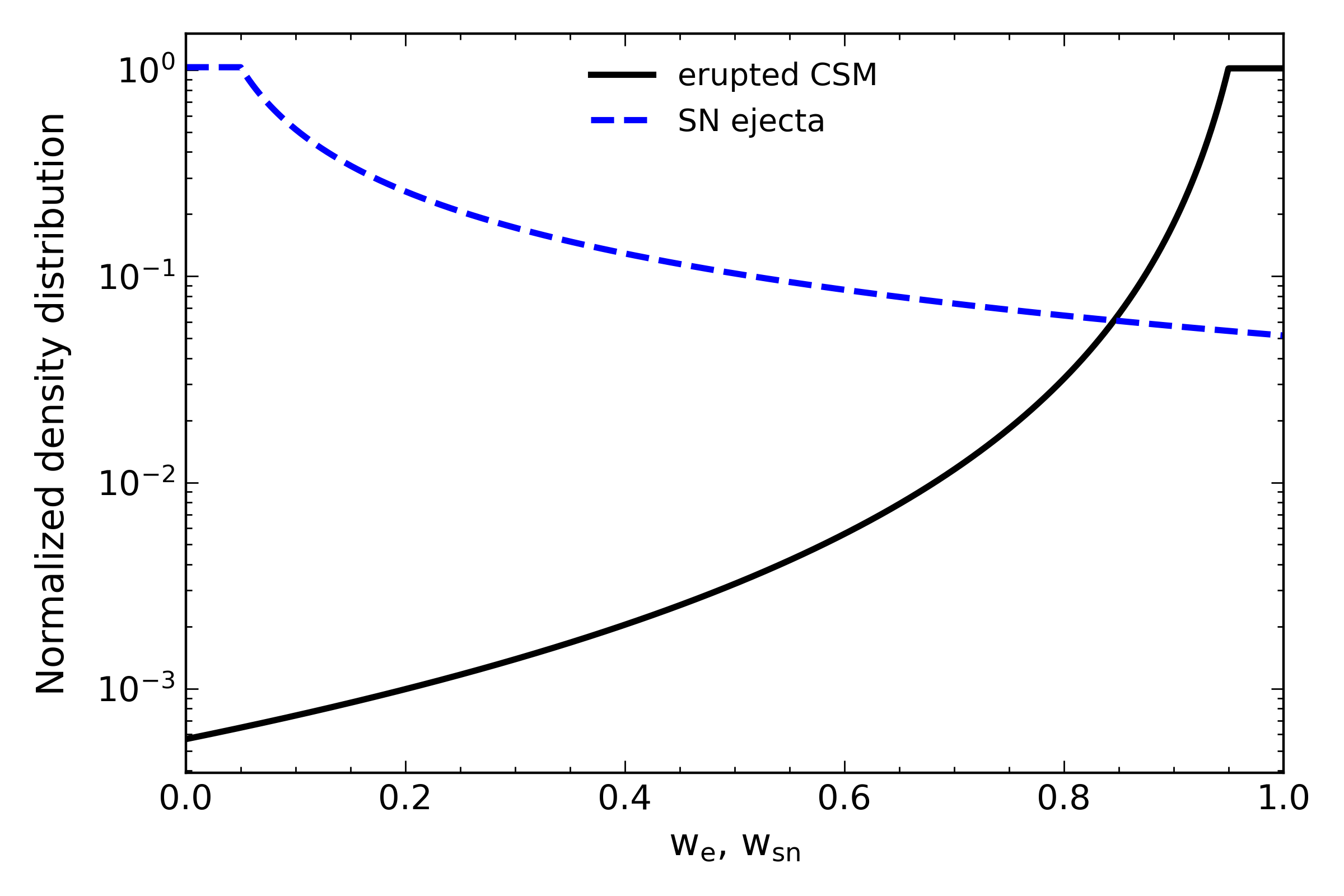}
    \caption{The normalized density distributions of ejected material during the eruption and the supernova ejecta are expressed as a function of $w_{\textrm{e}}=r/R_{\textrm{e}}$ and $w_{\textrm{sn}}=r/R_{\textrm{sn}}$, respectively. Note that the CSM follows a shell-like density structure.}
    \label{rho-e}
\end{figure}
We assume that the initial distribution for the ejecta density is given by:

\begin{equation}\label{erup_density_c}
    \rho_{\textrm{e}}(r) = \frac{M_{\textrm{e}}}{R_{\textrm{e}}^3}f_{\textrm{e}}(w_{\textrm{e}}),
\end{equation}
where $f_{\textrm{e}}(w_{\textrm{e}})$ is a structure function:
\begin{equation}\label{erup_density}
  f_{\textrm{e}}(w_{\textrm{e}})=\begin{cases}
         f_{k}(1-w_{\textrm{e}})^{- k} , \quad &\text{if} \  0\leq  w_{\textrm{e}} \leq w_{\textrm{sh,e}}, \\
        f_{0,{\textrm{e}}}, &\text{if} \  w_{\textrm{sh,e}} \leq w_{\textrm{e}} \leq 1, \\
       \end{cases}
\end{equation}
with a power-law index $k>0$,    $w_{\textrm{e}} = {r}/{R_{\textrm{e}}}$, and $w_{\textrm{sh,e}} = {R_{\textrm{sh,e}}}/{R_{\textrm{e}}}$.

Here, $r$, $R_{\textrm{sh,e}}$, and $R_{\textrm{e}}$ are the distance from the eruption site, and the inner and outer radii of the homogeneous eruption layer, respectively. The terms $f_{0,{\textrm{e}}}$ and $f_{k}$ are constants set by the mass and energy conservation (see Appendix \ref{AppendixA}). 

Fig. \ref{rho-e} presents the ejecta density distribution for $R_{\textrm{e}}=0.06$ pc, $k=2.5$ and $w_{\textrm{sh,e}}=0.95$, where one can note that density increases with $r$ up to a defined radius $R_{\textrm{sh,e}}$, followed by an uniform density outer shell between $R_{\textrm{sh,e}}$ and $R_{\textrm{e}}$. This approximation for the density distribution is designed to match a thin, hollow CSM shell mass distribution, similar to that observed in the Homunculus Nebula \citep[e.g.,][]{Smith2006,Smith2017,Smithetal2018,Steffenetal2014}, where the value of $w_{\textrm{sh,e}}=0.95$ corresponds to a shell thickness of $5\%$ of  $R_{\textrm{e}}$.

Note that similar density profiles as the one given by equations \eqref{erup_density_c}-\eqref{erup_density} have been used before to describe the initial conditions of other explosive events, such as SN remnants \citep[e.g.][]{TrueloveandMckee1999,TangChevalier2017}.

In order to model the bipolar morphology of the erupted CSM, we assume a latitude-dependent expansion velocity \citep{Smith2006}:
\begin{equation}
    v_{\textrm{e}} = v_{\textrm{r,e}}  \ F_{\varphi},
\end{equation}

where the radial velocity $v_{\textrm{r,e}}$ follows a Hubble-like expansion:
\begin{equation}\label{erup_radial_vel}
    v_{\textrm{r,e}}=\frac{v_{\textrm{max,e}}}{R_\textrm{e}}r,
\end{equation}
and
\begin{equation}\label{Func_F}
      F_{\varphi} = 1- \alpha\left( \frac{1-e^{-2\beta\sin^2{\varphi}}}{1-e^{-2\beta}} \right)
\end{equation}

is the function introduced by \cite{Franketal1995} to describe the angular dependence of the expanding ejecta \citep[see also][]{Blondin1995}. The parameter $\alpha$ controls the pole-to-equator velocity contrast, and $\beta$ controls the shape of the erupted CSM and $\varphi$ is the azimuth angle.
We have fixed $\alpha$ and $\beta$ to 0.78 and 0.3, respectively, in order to obtain a bipolar CSM similar to the Homunculus \citep[][]{Gonzalezetal2004}.

The maximum expansion velocity of the eruption ejecta is derived from energy conservation as: 
\begin{equation}
    v_{\textrm{max,e}} = \left[ \left( \frac{2E_{\textrm{k,e}}}{I_eM_{\textrm{e}}}\right)  \frac{1}{f_{k} \lambda_{v} + \frac{f_{0,\textrm{e}}}{5}(1-w_{\textrm{sh,e}}^{5})} \right]^{1/2},
\end{equation}
where
\begin{equation}
 I_e=2\pi\int_0^{\pi} F_{\varphi}^2 \sin (\varphi) d \varphi,
\end{equation}
and $\lambda_{v}$ is a constant dependent on the initial conditions (see Appendix \ref{AppendixA}).

\vspace{2mm}

In this setup, we assume that the stellar wind prior to the eruption is spherical, and the bipolar shape results from the eruption itself. However, in Appendix \ref{AppendixB}, we have also explored the opposite scenario, where the stellar wind is aspherical and the eruption is inserted with spherical symmetry. We demonstrate that both cases lead to similar CSM morphologies by the time the supernova explosion occurs.

Since we adopted an eruption ejecta mass of $M_{\textrm{e}}=25$ M$_\odot$, with a dust-to-gas mass ratio $\sim 0.01$  \citep{Smith&Ferland2007}, the ejected dust mass from the eruption is $M_{\textrm{dust}}=0.25$ M$_\odot$ in all our models. Furthermore, we assume that the kinetic energy of the eruptions is $E_{\textrm{k,e}}=10^{50}$ erg \citep{Smithetal2018, Smith&Andrews2020} in all cases. Finally, we consider only one eruptive event, acknowledging that massive stars may undergo multiple eruptions throughout their final stages of evolution \citep{Kochanek2011}.
\begin{table*}
\begin{center}
\caption{ Initial Conditions}
\begin{threeparttable}
\begin{tabular}{l c c c c l c c c c }
    \hline
    \hline
     Case   &$M_{ZAMS}$& $R_{\textrm{w}}$ & $R_{\textrm{e}}$ &$R_{\textrm{sn}}$  &$M_{\textrm{sn}}$ &$E_{\textrm{sn}}$ &$^{56}$Ni mass& Min. Res. & Max. Res.   \\  
            &\scriptsize (M$_{\odot}$)& \scriptsize(pc) &\scriptsize (pc)&\scriptsize (pc)  & \scriptsize (M$_{\odot}$)&\scriptsize (erg) &\scriptsize (M$_{\odot}$)&\scriptsize (pc) &\scriptsize (pc)   \\
 & & & & & & && &  \\ \hline
 
     $S_{200}$                 &60 & 0.04  & 0.06   & 0.04    &10 &$10^{51}$&-& $1.5\times 10^{-2}$& $3.9\times 10^{-3}$  \\
     $S_{200-\textrm{Test}}$   &60& 0.04  & 0.06   & 0.04    &10 &$10^{51}$ &-& $1.5\times 10^{-2}$& $7.8\times 10^{-3}$  \\ 
     $B_{200}$                 &60& 0.04  & 0.06   & 0.04    &10 &$10^{51}$ &-& $7.8\times 10^{-3}$& $1.9\times 10^{-3}$  \\
     $B_{200-\textrm{Test}}$    &60& 0.04  & 0.06   & 0.04    &10 &$10^{51}$ &-& $7.8\times 10^{-3}$& $7.8\times 10^{-3}$  \\  
     $B_{12-10}$ &60& 0.008 & 0.0012 & 0.0015  &10 &$10^{51}$ &-& $1.9\times 10^{-4}$& $4.8\times 10^{-5}$   \\   
     $B_{12-10-\textrm{Test}}$  &60& 0.008 & 0.0012 & 0.0015  &10 &$10^{51}$ &-& $1.9\times 10^{-4}$& $9.7\times 10^{-5}$  \\
     $B_{12-\textrm{56Ni}}$    &60& 0.008 & 0.0012 & 0.0015  &10 &$10^{51}$ &0.032& $1.9\times 10^{-4}$& $4.8\times 10^{-5}$  \\     
     $B_{12-15}$ &45& 0.008 & 0.0012 & 0.0015  &15 &$5 \times 10^{51}$ &0.9& $1.9\times 10^{-4}$& $4.8\times 10^{-5}$  \\  
     $B_{12-20}$ &50& 0.008 & 0.0012 & 0.0015  &20 &$5 \times 10^{51}$ &2.5& $1.9\times 10^{-4}$& $4.8\times 10^{-5}$  \\ \hline  
\end{tabular}
\begin{tablenotes}
      \footnotesize
      \item \textbf{Notes.} 
For each case, we assume for the stellar wind a mass loss rate $\dot{M}_{w} = 10^{-3}$ M$_\odot$ yr$^{-1}$, and a terminal velocity $v_{\infty}= 250$ km s$^{-1}$. The eruption mass and energy are 25 M$_{\odot}$ and $10^{50}$ erg, respectively, with a dust mass of 0.25 M$_{\odot}$. 
From left to right, the table presents the ZAMS mass of the progenitor, the insertion radius of the stellar wind, eruption and SN remnant, the mass of $^{56}$Ni, along with both the minimum and maximum resolutions for each case.
\end{tablenotes}
\end{threeparttable}
\label{table1}
\end{center}
\end{table*}

\subsection{Supernova remnant evolution}\label{sec_sn}

In order to model the SN remnant evolution, we consider as input parameters the total kinetic energy $E_{\textrm{k,sn}}$, the maximum SN ejecta expansion velocity $v_{\textrm{max,sn}}$ and the total ejected mass $M_{\textrm{sn}}$. The insertion of supernova remnant is performed analogously to the insertion of the erupted CSM. Thus, following \citet{TangChevalier2017} and \citet{TrueloveandMckee1999} \citep[see also][]{Chevalier1982}, the density distribution is given by:
  \begin{equation}
    \rho_{\textrm{sn}}(r) = \frac{M_{\textrm{sn}}}{R_{\textrm{sn}}^3}f_{\textrm{sn}}(w_{\textrm{sn}}),
  \end{equation}

\begin{equation}
  f_{\textrm{sn}}(w_{\textrm{sn}})=\begin{cases}
        f_{0,{\textrm{sn}}}, \quad &\text{if} \  0\leq  w_{\textrm{sn}} \leq w_{\textrm{c,sn}}, \\
        f_{m}w_{\textrm{sn}}^{-m}  &\text{if} \  w_{\textrm{c,sn}} \leq w_{\textrm{sn}} \leq 1, \\
       \end{cases}
  \end{equation}
where $R_{\textrm{sn}}$ is the SNR insertion radius, $w_\textrm{sn} = r/R_{\textrm{sn}}$, and $w_{\textrm{c,sn}} = {R_{\textrm{c,sn}}}/{R_{\textrm{sn}}}$.

We note that SNe follow a decreasing power-law density distribution  as depicted in Fig. \ref{rho-e} for  $R_{\textrm{sn}}=0.04$ pc, and $w_{\textrm{c,sn}}=0.05$.
The index $m$ of the power-law envelope region has been chosen to be equal to 1.

The initial radial velocity of the SN ejecta follows a profile analogous to the one given by equation \ref{erup_radial_vel} but we also assume that the SN ejecta is initially spherically symmetric. In such a case, energy conservation defines the maximum velocity of the expanding SN ejecta:
\begin{equation}
    v_{\textrm{max,sn}} = \left[2w_{\textrm{c,sn}}^{-2}\left( \frac{5-m}{3-m}\right)  \left( \frac{w_{\textrm{c,sn}}^{m-3}-m/3}{w_{\textrm{c,sn}}^{m-5}-m/5}\right) \frac{E_{\textrm{k,sn}}}{M_{\textrm{sn}}}\right]^{1/2}.
\end{equation}

\vspace{2mm}
We assume four cases for hydrogen--rich core-collapse SN explosions. In the first one ($B_{12-10}$), a standard ejecta mass and kinetic energy of $10 \ \textrm{M}_\odot$ and $10^{51}$ erg \citep{vanMarleetal2010, Dessartetal2016, Matsumotoetal2024}, respectively, are considered. In the second case ($B_{12-\textrm{56Ni}}$), we use the same parameters as in the $B_{12-10}$ case, but with the SN ejecta set at an initial temperature of $10^6$ K aimed to mimic the effect of radioactive decay heating (see Appendix \ref{AppendixC}). In the third case ($B_{12-15}$), a more massive and energetic explosion is assumed, releasing $15 \ \textrm{M}_\odot$ of gas with a kinetic energy of $5 \times 10^{51}$ erg. This is consistent with SN iPTF14hls, whose estimated energy is between $10^{51}-10^{52}$ erg, and which has an ejecta mass greater than $10 \ \textrm{M}_\odot$ \citep{Andrews&Smith2018, Wangetal2022}.
For the fourth case, an even more massive SN ejecta is assumed, releasing $20 \ \textrm{M}_\odot$ of gas with a kinetic energy of $5 \times 10^{51}$ erg \citep[e.g. the SN 2006gy,][]{Smithetal2010,Moriyaetal2013}.

In the latter cases $B_{12-15}$ and $B_{12-20}$, we assumed $M_{\textrm{Ni}}\sim 0.9 M_{\odot}$ and $2.5 M_{\odot}$, respectively \citep{Wangetal2022, Moriyaetal2013}. Thus, the corresponding energies 
released by the radioactive decay of $^{56}$Ni
(see equation \ref{E56Ni} in Appendix \ref{AppendixC}) result into ejecta temperatures $> 10^7$ K.
In the three cases considering the radioactive decay heating, we disabled radiative cooling for the first 200 days of the SN remnant's evolution, allowing it to cool down solely through adiabatic expansion \citep[although cooling from processes like CO rovibrational transitions should be at play][]{Onoetal2024}.

As we have done for the stellar wind, we refrain from injecting any dust along with the SN ejecta. This allows us to focus on individually tracking the dust originating from the erupted CSM. In doing so, we aim to better understand the fate of CSM-formed dust grains, recognizing the complexity that would arise from attempting to trace dust grains from multiple sources.

\begin{figure*}
    \centering
    \captionsetup{margin=25pt}
    \captionsetup{skip=1pt}   \includegraphics[width=0.85\textwidth]{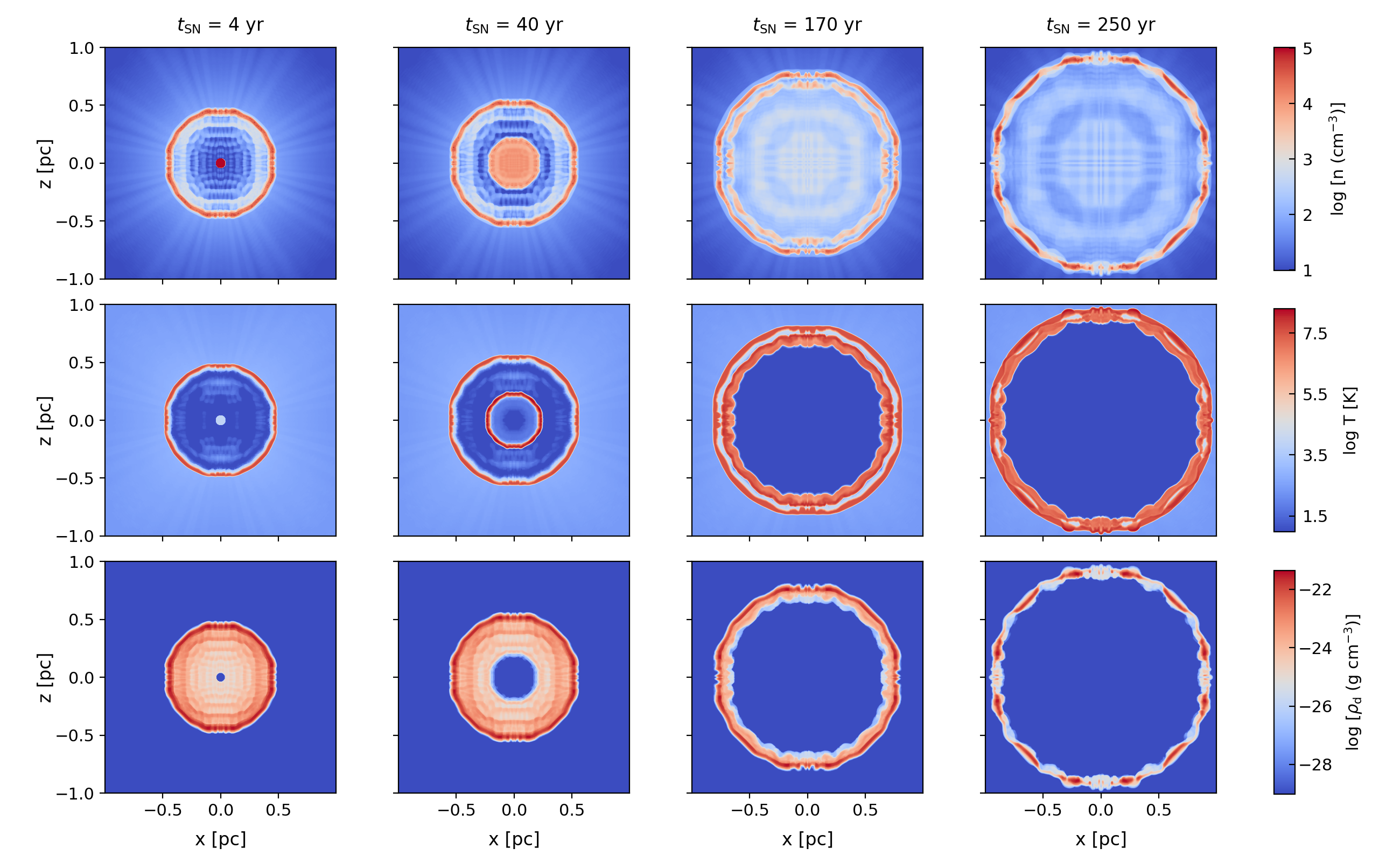}
    \caption{The evolution of the supernova remnant within the erupted circumstellar medium for $S_{200}$, with each column showing  snapshots at $t_{\textrm{SN}}=4, 40, 170, \ \textrm{and} \ 250$ yr after the supernova explosion, respectively. From top to bottom, each row displays, in log scale, the number gas density, the gas temperature, and the dust mass density.}
    \label{CaseS200}
\end{figure*}

\begin{figure*}
    \centering    
    \captionsetup{margin=25pt}
    \captionsetup{skip=1pt}    \includegraphics[width=0.87\textwidth]{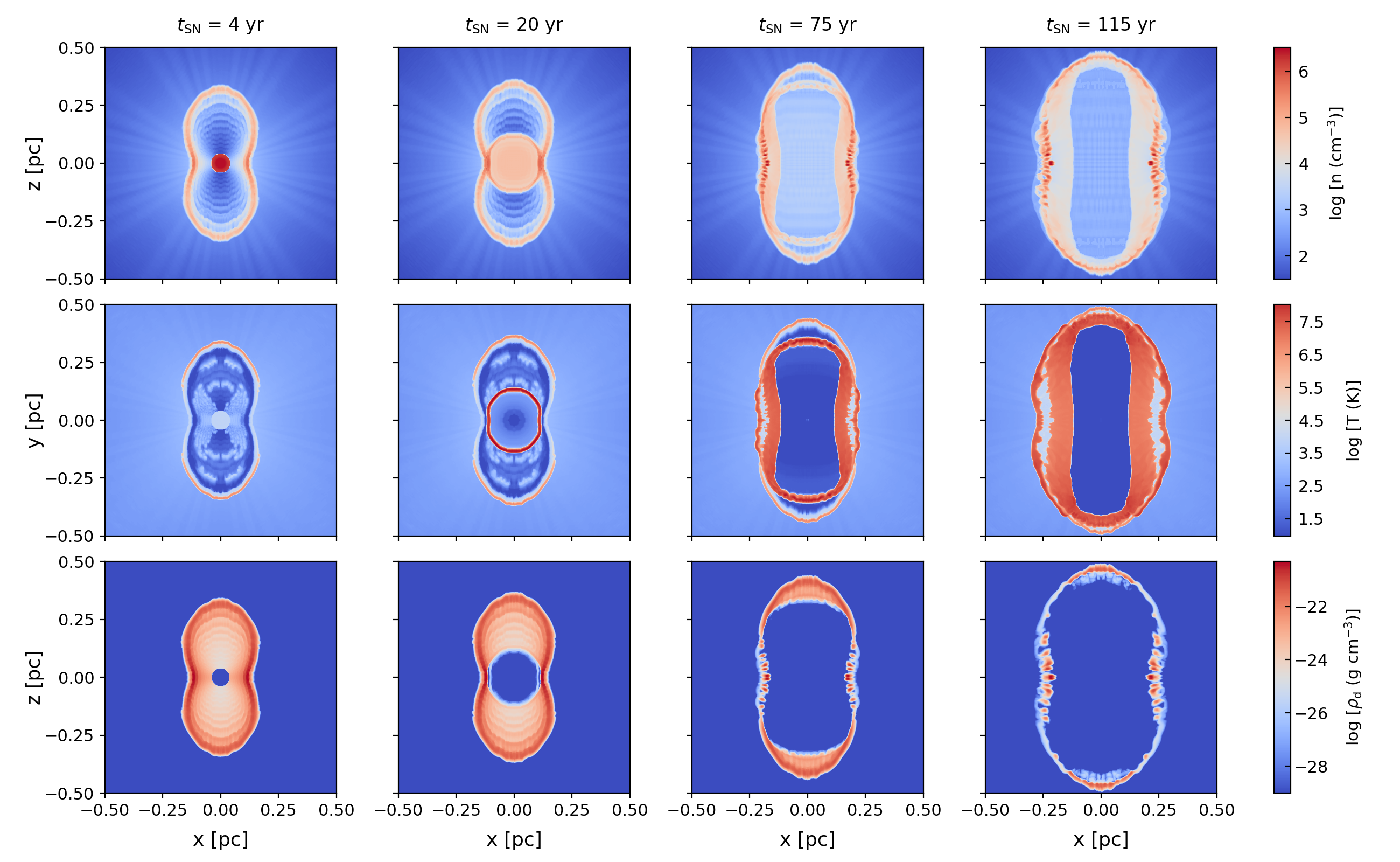}
    \caption{Same as Fig. \ref{CaseS200} but for $B_{200}$, with snapshots at $t_{\textrm{SN}}=4, 20, 75, \ \textrm{and} \ 115$ yr,  respectively. 
    Additional one-dimensional density profiles along the x and z axes are provided in Appendix \ref{AppendixD}.}
    
    \label{CaseB200}
\end{figure*}

\subsection{Dust injection and destruction} \label{sec_dust}

The injection and destruction of dust in the simulations was carried out with the \textsc{Cinder} module, that also allows to calculate, on-the-fly, the rate of the thermal sputtering within specific grid cells.
\textsc{Cinder} incorporates an initial grain size distribution with 10 bins logarithmically spaced, and the dust is traced by means of the dust-to-gas mass ratio as an advecting mass scalar. The grain sizes follow a log-normal distribution of the form $\sim a^{-1}\exp \{-0.5[\ln(a/a_0)/\sigma]^2 \}$, where $a$ is the grain size, $a_{0}=0.1$, $\sigma=0.7$, and $a_{min}=0.005 \ \mu \textrm{m}$, $a_{max}=0.5 \ \mu \textrm{m}$ are the lower and upper limits of the grain size, respectively \citep[see][]{MartinezGonzalezetal2018,MartinezGonzalezetal2021}.

\begin{figure*}
    \centering 
    \captionsetup{skip=1pt}
    \includegraphics[width=0.90\textwidth]{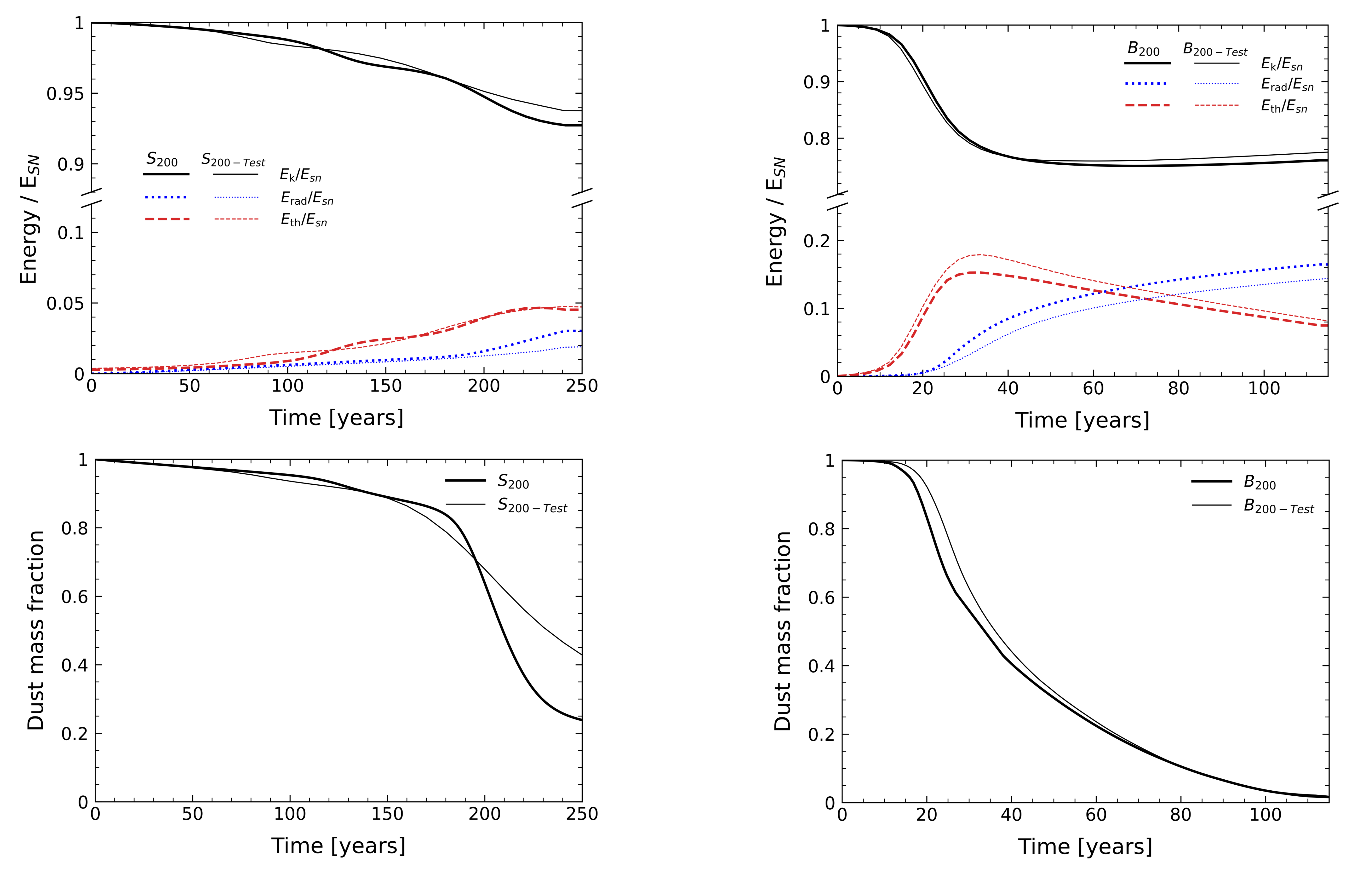}
    \caption{Upper panels: the supernova remnant (only gas) kinetic (solid), thermal (dotted), and radiated (dashed) energies as a function of time for $S_{200}$ (left) and $B_{200}$ (right).
    Lower panels: the evolution of the dust mass, normalized to that ejected by the eruption. For all panels, thicker lines correspond to $S_{200}$ and $S_{200}$ while thinner lines to $B_{200-\textrm{Test}}$ and $S_{200-\textrm{Test}}$, respectively. }
    \label{dust-energy1}
\end{figure*}

\subsection{Initial conditions}
Our simulations initialize with a spherical stellar wind that undergoes evolution within an ambient medium of number density $n = 1 \ \textrm{cm}^{-3}$.
The presence of this wind sets the conditions for the subsequent evolution of the erupted CSM. The latter is simulated considering different cases with spherical and bipolar (similar to the Homunculus nebula) geometries. Such an eruption leads to produced $M_{\textrm{dust}}=0.25$ M$_\odot$ of dust. Finally, the simulations take into account the effects of a spherical SN explosion that interacts with the pre-existing CSM created by the eruption and the stellar wind. Additionally, the growth of dust grains is also included using the capabilities of the \textsc{Cinder} module, as described in \cite{MartinezGonzalezetal2022}.
The simulations incorporate white noise, i.e., random initial density perturbations.

We explored three distinct scenarios based on the shape of the erupted CSM, and the eruption-SN gap. For the first and second scenarios, the SN explosion is considered to occur $200$ yr after the CSM ejection as a result of the eruptive episode. This time interval aligns with cases such as SN 2004dk and SN 2007od, whose estimated CSM-SN delay time is >100 years \citep{Andrewsetal2010, Mauerhanetal2018}, and it is also comparable with the time elapsed from the Great Eruption of $\eta$ Car to the present day.
These two scenarios consider a spherically symmetric CSM (case $S_{200}$), and a bipolar-shaped CSM (case $B_{200}$), respectively. 

In certain instances, the eruption may occur just a few years prior to the SN explosion, as exemplified by the SNe 2006qq, 2006Y, 2017hcc, 2018zd, and 2020tlf, whose estimated evolution time is within the range of $6$\textendash$12$ years \citep{Taddiaetal2013, Smith&Andrews2020, Hiramatsuetal2021a, Hiramatsuetal2021b, Chugai&Utrobin2022}.
Therefore (in cases $B_{12-10}$, $B_{12-56Ni}$, $B_{12-15}$, and $B_{12-20}$), the SN explosion occurs $12$ yr after a bipolar CSM ejection.

Table \ref{table1} presents the initial parameters adopted for the stellar wind, the erupted CSM and the SN remnant for each case.

The simulations in the $S_{200}$ and $B_{200}$ cases take place in a cubic computational domain where the origin of the Cartesian coordinate system lies at the center, and whose size is $(2 \ \textrm{pc})^3$, and $(1 \ \textrm{pc}) ^3$ for cases $S_{200}$ and $B_{200}$, respectively. In the $B_{12}$ cases, the simulations were performed in a single octant of the computational domain with size $(0.05 \ \textrm{pc})^3$.

A minimum refinement level of 5 and maximum of 7 \footnote{At a fixed cell size, refinement level 5 corresponds to an equivalent uniform grid of $(128)^3$ cells, and refinement level 7 corresponds to $(512)^3$ cells. Similarly, at refinement level 6, there are $(256)^3$ cells on the equivalent uniform grid, and at refinement level 8, there are $(1024)^3$ cells.} are considered for $S_{200}$ and $B_{200}$, and 6 and 8 for $B_{12-10}$, $B_{12-56Ni}$, $B_{12-15}$, and $B_{12-20}$. We set the minimum and maximum spatial resolutions shown in Table \ref{table1}. The external boundary conditions are set to outflow in cases $S_{200}$ and $B_{200}$, while in the simulations conducted within a single octant, reflecting boundary conditions were implemented on the planes $x=0$, $y=0$, and $z=0$. 
Low-resolution convergence tests were also conducted for the $S_{200}$, $B_{200}$ and $B_{12-10}$ cases, with the corresponding minimum and maximum resolutions listed in Table \ref{table1}.

\section{Results} \label{results}

\begin{figure*}
    \centering
    \includegraphics[width=0.8\textwidth]{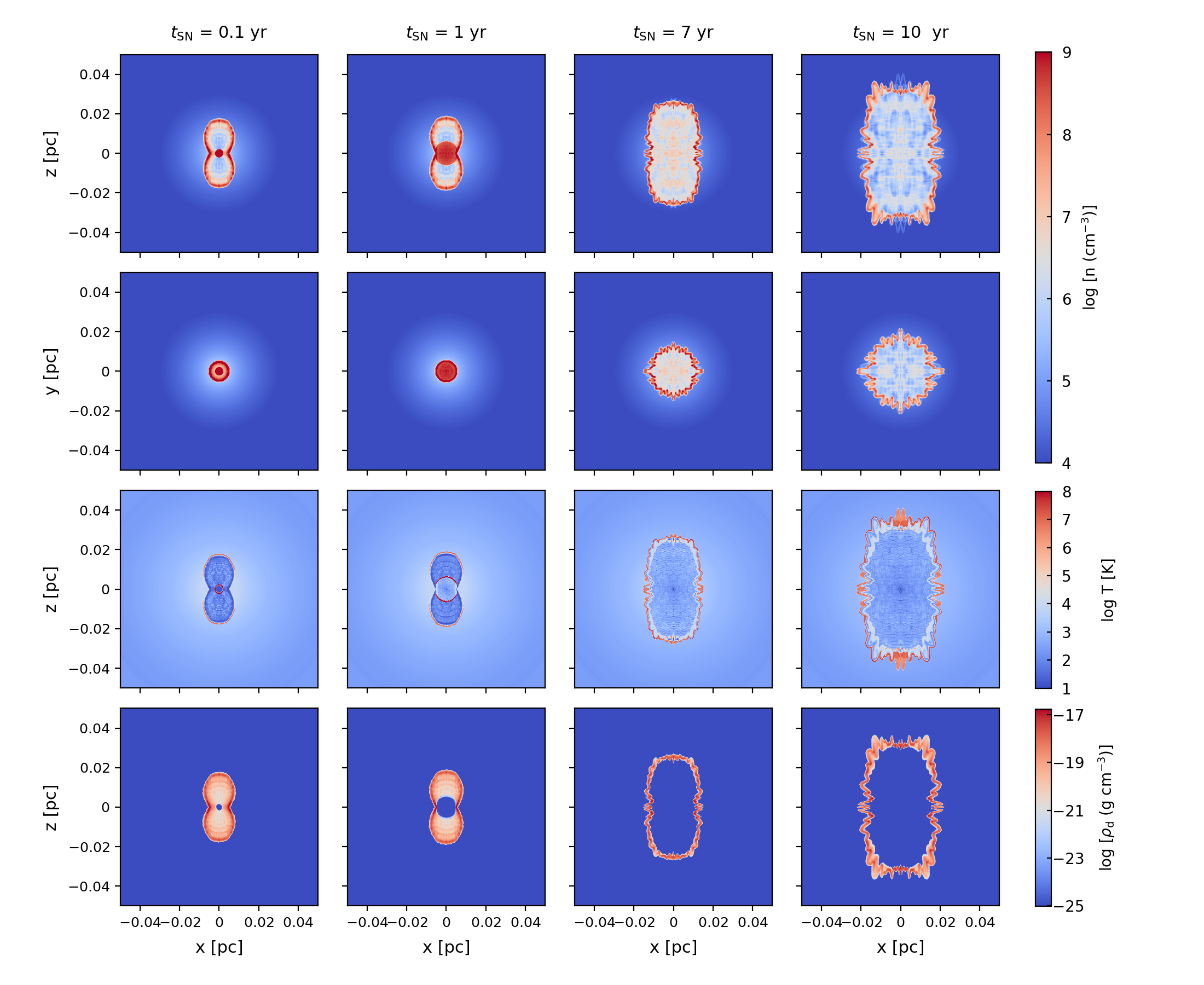}
    \captionsetup{skip=1pt}
    \caption{Same as Fig. \ref{CaseB200} but for $B_{12-10}$, with snapshots at $t_{\textrm{SN}}=0.1, 1, 7 \ \textrm{and} \ 10$ yr after the SN explosion, respectively. Note in the last column that several {\em blisters} of shock-heated gas already extend beyond the CSM density distribution. Additional one-dimensional density profiles along the x and z axes are provided in Appendix \ref{AppendixD}.}
    \label{CaseB12}
\end{figure*}

\begin{figure*}
    \centering
    \includegraphics[width=0.8\textwidth]{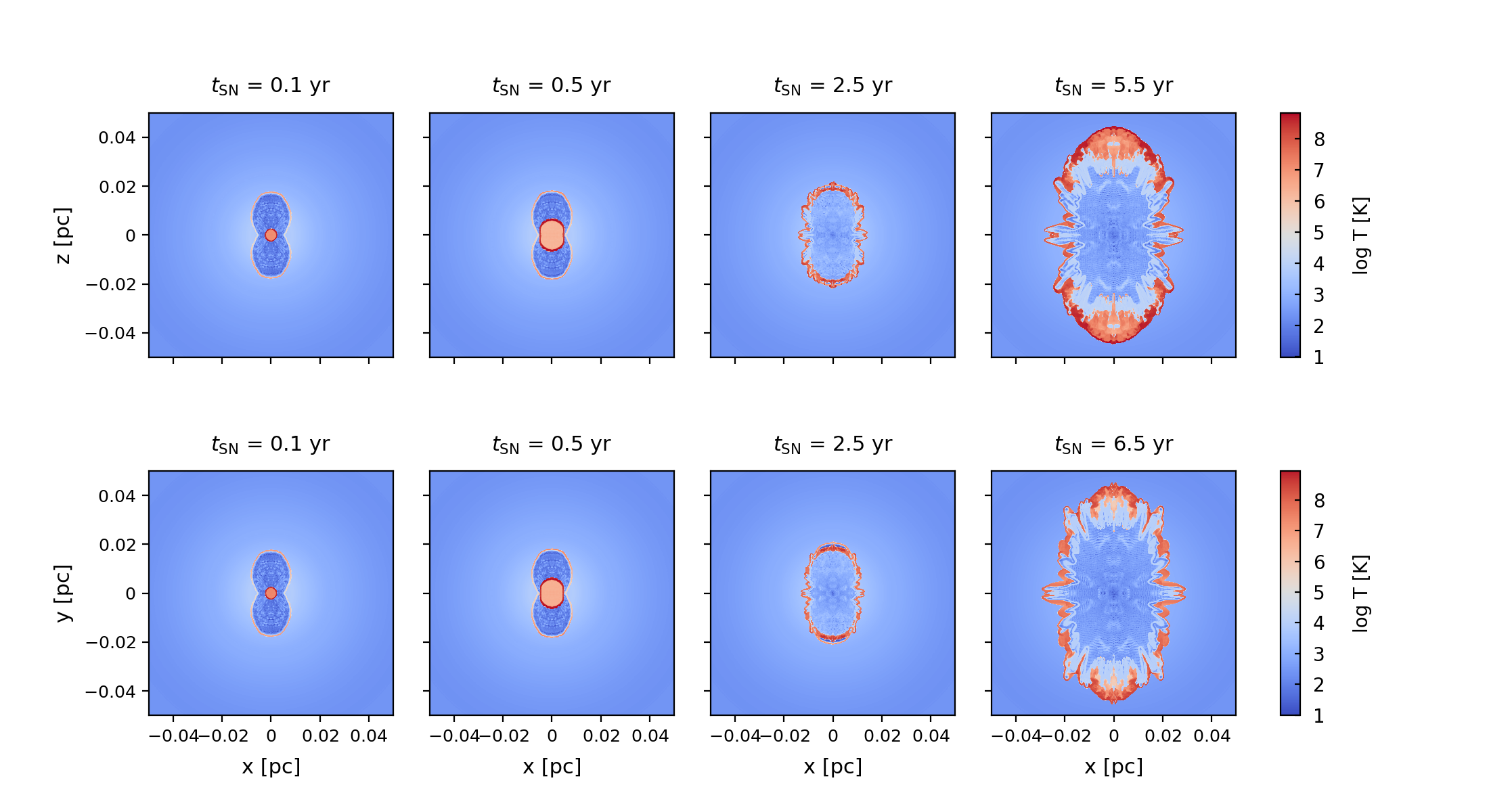}
    \captionsetup{skip=1pt}
    \caption{Gas temperature distribution slices in the $x$\textendash$z$ plane, with the first row showing results for $B_{12-15}$ and the second row for $B_{12-20}$.}
    \label{CaseB12-15-20}
\end{figure*}

Figures \ref{CaseS200}, \ref{CaseB200}, and \ref{CaseB12}, present $x$\textendash$z$ slices of the logarithm of the number gas density, temperature, and dust mass density of four snapshots for the $S_{200}$, $B_{200}$, and $B_{12-10}$, respectively. Similarly, Fig. \ref{CaseB12-15-20} present the logarithm of the gas temperature only. For the $B_{12-10}$, $B_{12-15}$ and $B_{12-20}$ cases, Figs. \ref{CaseB12} and \ref{CaseB12-15-20} show the plots mirrored along the $x$ and $z$ axes to illustrate the simulations as if they were conducted across the full domain instead of a single octant. Appendix \ref{AppendixD} also presents 1D density profiles along the x and z axis for $B_{200}$ and $B_{12-10}$, for the same post-SN times as in Figs. \ref{CaseB200} and \ref{CaseB12}, respectively.

\begin{figure}
    \centering
    \captionsetup{skip=1pt}
    \includegraphics[width=0.48\textwidth]{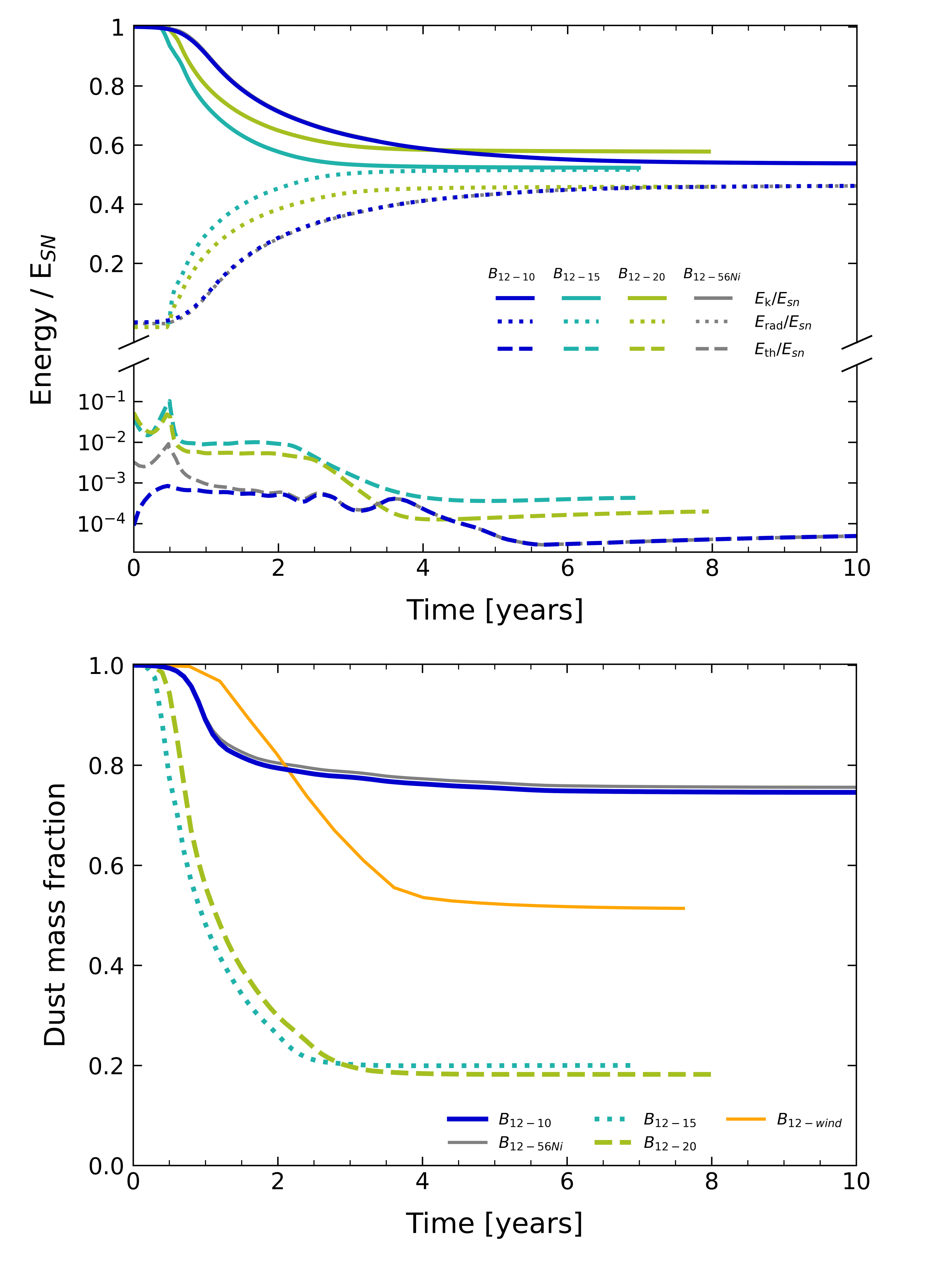}
    \caption{Same as Fig. \ref{dust-energy1} but for $B_{12-10}$, $B_{12-\textrm{56Ni}}$, $B_{12-15}$, $B_{12-20}$, and $B_{12-wind}$ (see Appendix \ref{AppendixB}).Note that while the curves corresponding to cases $B_{12-10}$ and $B_{12-56Ni}$ are presented up to 10 years post-supernova, the ones corresponding to $B_{12-15}$, $B_{12-20}$, and $B_{12-wind}$ are presented only up to 7--8 years, as the simulations have reached the boundaries of the computational domain, but the most relevant phase of their evolution has already taken place.}
    \label{dust-energy2}
\end{figure}

Figs. \ref{dust-energy1} and \ref{dust-energy2} show the evolution of the SN kinetic, thermal and radiated energies, and the dust mass fraction after the SN explosion for all the models listed in Table \ref{table1}.

\subsection{\texorpdfstring{200-year eruption-SN delay}{200-year eruption-SN delay}} \label{sec_200}

The $S_{200}$ (spherical) and $B_{200}$ (bipolar) cases are characterized by a distinct geometry of the CSM created by the eruption. In both cases, the stellar eruption occurs 200 years before the SN explosion. The top left panels in Figs. \ref{CaseS200} and \ref{CaseB200} show the instant a few years after the SN explosion ($t_{\textrm{SN}}=4$ yr), when the shell-like CSM have reached maximum number densities of $\sim 10^{4.5}$ cm$^{-3}$ and $\sim 10^{6}$ cm$^{-3}$ for $S_{200}$ and $B_{200}$, respectively. These densities
are lower than those found in the outer shell of the Homunculus
Nebula, which average approximately $10^7$ cm$^{-3}$ \citep{Smith2006}.

For case $S_{200}$, the SN forward shock initially grows almost in free expansion given the relatively low density cavity created by the preceding stellar eruption. Note that the kinetic energy (solid line in the top left panel of Fig. \ref{dust-energy1}) decreases slightly and the thermal energy (dashed line) is very small during these early stages ($t_{\textrm{SN}} \lesssim 170$ yr), thus indicating that indeed the forward shock has processed only a small fraction of the circumstellar gas. On the other hand, the high shock velocity and post-shock temperatures ($T \sim 10^{7}$ K, see the middle row of Fig. \ref{CaseS200}) lead to an efficient destruction of the CSM dust located in those inner regions, as shown in the bottom row of Fig. \ref{CaseS200}. However, the dust mass fraction decreases just $\sim 10 \%$ (see the bottom panel of Fig. \ref{dust-energy1}) as only a small amount of dust is located within the cavity. Once the forward shock reaches the densest region of the CSM at $t_{\textrm{SN}} \sim 170$ yr, the shock processes the largest fraction of dust in the shell, thus increasing considerably the amount of dust destroyed, as shown by the steeper decline on the dust mass fraction in the bottom left panel of Fig. \ref{dust-energy1}. After 250 years post-explosion, around 25\% of the dust mass ($\sim 0.06$ M$_{\odot}$) produced in the erupted CSM still remains. Finally, it is also interesting to note that the middle row in Fig. \ref{CaseS200} shows that up to the simulated time, the SN reverse shock has not started moving inwards, which may be due to the still very low thermal energy (and thus thermal pressure) behind the forward shock. For this case the impact of gas and dust cooling is negligible for the simulated time, see $E_{\textrm{rad}}$ (dotted line) in the top left panel of Fig. \ref{dust-energy1}. 

\vspace{0.2cm}
The SN remnant evolution within the bipolar CSM (case $B_{200}$) proceeds somewhat differently. First, in this case, the cavity density is larger than in $S_{200}$ (see the top row in Fig. \ref{CaseB200}). Also, at $t_{\textrm{SN}}=20$ yr the SN forward shock collides with the equatorial section of the CSM shell. Hence, the thermalization of the SN energy proceeds faster. This is shown in the top right panel of Fig. \ref{dust-energy1}, where the SN kinetic energy drops rapidly while being transformed into kinetic and thermal energy of the shocked gas. Furthermore, this is also supported by the fact that, unlike $S_{200}$, in this case the SN reverse shock starts moving towards the center (see the middle row in Fig. \ref{CaseB200}) very early in the evolution of the remnant ($t_{\textrm{SN}} \sim 70-100$ yr). 
It becomes increasingly evident that the SN remnant undergoes deformation, acquiring a configuration that closely resembles the bipolar shape of the CSM, as depicted in Fig. \ref{CaseB200}. This suggests that the SN remnant is constrained from expanding freely in the direction of those sections of the CSM, due to the high equatorial densities.

The lower panels of Fig. \ref{dust-energy1} show that the dust destruction in $B_{200}$ occurs more efficiently as compared to $S_{200}$.
As a result, in the bipolar case only $\sim$ 2\% of  the CSM dust survives within $115$ years post-explosion. However, we argue that the main difference between $S_{200}$ and $B_{200}$ regarding the fate of the CSM dust is the time-scale of the destruction process, which is smaller in $B_{200}$ only because the forward shock encounters the bulk of the CSM dust (located in the dense shell) faster than in $S_{200}$. Ultimately, however, the eruption/SN composite remnant will keep expanding, and the destruction of the CSM dust grains will continue in both cases. Given the similarities to the $B_{200}$ case, it is expected that in a circumstellar nebula like the Homunculus, all the dust produced after the eruption would be destroyed if the progenitor exploded as a SN.

Note that as soon as the forward shock starts its interaction with the CSM shell, the radiated energy increases considerably (dotted line in the top right panel of Fig. \ref{dust-energy1}). In fact, the remnant thermal energy reaches a peak value of $\sim 0.15 E_{\textrm{SN}} $ at $t_{\textrm{SN}} \sim 25$ yr, and then starts to fall afterwards, which could delay the reverse shock in its motion towards the explosion center. This maximum for the gas thermal energy differs considerably with the one expected for an adiabatic SN remnant ($ \sim 0.7 E_{\textrm{SN}} $). Hence, although gas cooling is not sufficient to decrease the destruction rates of the CSM dust grains in this particular case ($B_{200}$), it could potentially hinder the ability of the SN forward shock to destroy the interstellar dust contained within the wind-driven shell located at a greater distance at later stages of its evolution \citep{MartinezGonzalezetal2019}. Finally, between $t_{\textrm{SN}}=50$ yr and $t_{\textrm{SN}}=115$ yr, one can observe a slight increase in the kinetic energy as the SN shock wave reaches and expands into the decreasing wind density, and therefore it accelerates \citep[e.g.][]{TenorioTagleetal2015, Jimenezetal2021}.

\vspace{0.2cm}
We conducted convergence tests at lower resolution for both $S_{200}$ and $B_{200}$ cases (namely, $S_{200-\textrm{Test}}$ and $B_{200-\textrm{Test}}$) using a maximum refinement level of 6, obtaining the resolutions shown in Table \ref{table1}. The results, depicted in Fig. \ref{dust-energy1} by thin lines, demonstrate an agreement within 1\% for the energy and within 2\% for the dust mass fraction in $S_{200}$, and similarly, an agreement within 1\% for the energy and 1\% for the dust mass fraction for $B_{200}$.

\subsection{\texorpdfstring{12-year eruption-SN delay}{12-year eruption-SN delay}} \label{sec_B12}

The $B_{12-10}$, $B_{12-\textrm{56Ni}}$, $B_{12-15}$ and $B_{12-20}$ models also consider a bipolar CSM but with a 12 years delay between the onset of the stellar eruption and the SN explosion.

Due to the short delay time, the CSM cavity is much denser as compared to $S_{200}$ and $B_{200}$, reaching densities $\sim 10^9$ cm$^{-3}$ mostly concentrated at equatorial latitudes (see the top row of Fig. \ref{CaseB12}). Thus, the shock wave undergoes a rapid transition to a radiative phase, therefore bypassing the Sedov-Taylor phase \citep{Panetal2013}.

For the $B_{12-10}$ case, the solid line in the bottom panel of Fig. \ref{dust-energy2} shows that although the dust grains begin to be processed soon, by the time the SN reaches the densest region of the CSM ($t_{\textrm{SN}}=1$ yr, second column in Fig. \ref{CaseB12}), dust destruction is limited, and as much as 75\% (0.19 M$_\odot$) is preserved. 

In the $B_{12-\textrm{56Ni}}$ case, the thermal, kinetic, and radiated energy curves nearly overlap with those of $B_{12-10}$, appearing as single curves in each plot. Here, about 76\% of the dust mass is conserved (thin solid grey line in Fig. \ref{dust-energy2}), slightly higher than in the $B_{12-10}$ case. This is because the SN shell has a slightly larger thickness (hence, it is less dense) due to the absence of radiative cooling, leading to less dust destruction. We note that as the SN ejecta becomes optically thin to $\gamma$-rays around 200 days after the explosion, the radioactive heating efficiency is expected to sharply decrease over time \citep{Matsumotoetal2024}.

Conversely, for $B_{12-15}$ and $B_{12-20}$, where the explosion energy is higher, the SN remnants expand more rapidly, reaching the equatorial region of the CSM shell in a shorter time, as shown in the second column of Fig. \ref{CaseB12-15-20}. Consequently, at $t_{\textrm{SN}} = 0.5$ yr, the stronger shock induces a more efficient dust destruction, resulting in preserving about 20\% of the dust mass (see the dashed and dotted curves in the lower panel of Fig. \ref{dust-energy2}).

In all $B_{12}$ cases, namely, $B_{12-10}$, $B_{12-\textrm{56Ni}}$, $B_{12-15}$, $B_{12-20}$ and $B_{12-\textrm{wind}}$ (see Appendix \ref{AppendixB}), stabilized dust mass fractions have been reached by the end of the simulations, suggesting prolonged survival of the remaining CSM dust grains. 

Note that in case $B_{12-10}$ the gas thermal energy remains orders of magnitude lower than the kinetic energy due to the strong gas and dust cooling, as shown by the dashed and dotted lines in Fig. \ref{dust-energy2}. For cases $B_{12-15}$, and $B_{12-20}$, the amount of gas thermal energy is higher, reaching up to $5\times10^{50}$ erg due to the injection of energy via \(^{56}\textrm{Ni}\) radioactive decay heating (see Appendix \ref{AppendixC}). The radiated energy corresponding to case $B_{12-10}$ reaches values of $>0.4E_{\textrm{SN}}$ ($=4 \times 10^{50}$ erg) at $t_{\textrm{SN}}=$ 10 yr, which is comparable with that obtained by \citet{vanMarleetal2010} for their two-dimensional B04 model with similar parameters (a $25$ M$_{\odot}$ CSM shell and a SN explosion injecting $10$ M$_{\odot}$ of gas) for a 2-year eruption-SN gap.
On the other hand, for $B_{12-15}$ and $B_{12-20}$, the cumulative radiated energy reaches approximately $E_{\textrm{rad}} \sim 0.5E_{\textrm{SN}}$ (equivalent to $2.5 \times 10^{51}$ erg) after $t_{\textrm{SN}} = 7$ and $t_{\textrm{SN}} = 8$ years, respectively. Therefore, it is expected that this energy should be emitted as infrared radiation by shock-heated CSM dust grains 
\citep{Chevalier&Fransson2017,Tartagliaetal2020,Dweketal2021}.

As can be observed in the top and middle panels in Fig. \ref{CaseB12}, the SN shock wave manages to overrun the whole erupted CSM, leaving a cold region behind. According to \cite{TenorioTagleetal1990}, the mass of a an encompassing dense shell must be approximately 40 times greater than that of the SN ejecta to effectively prevent its overrunning by the SN blast wave. However, in the current scenario, this criterion is far from being fulfilled, as the mass of the erupted CSM is only 2.5 times the SN ejecta mass.

Counter-intuitively, our results reveal that for a given SN energy, a shorter delay time between an eruptive episode and the subsequent SN explosion results in a larger fraction of surviving dust mass. This unexpected outcome is attributed to the enhanced radiative cooling as the dusty CSM is still very dense (reaching densities as high as a few $10^9$ cm$^{-3}$) at its early stages. This result emphasizes the crucial role of the time interval in shaping the fate of dust grains generated after stellar eruptions.
It also shows that pre-existing dust grains in the surroundings of SNRs with late CSM interactions have a greater chance of survival.

Finally, the growth of {\em blisters} of hot gas is observed on the surface of the eruption/SN composite remnant, as depicted in the middle row of Fig. \ref{CaseB12} and in Fig. \ref{CaseB12-15-20}. They are created by the interaction of the SN shock that has passed through the CSM with the slow stellar wind.
This is similar to the scenario described by \citet{Pittard2013}, except that the hot gas is not escaping from the interior of the CSM shell, but is generated by the shock on the shell surface.
In our model, these features are resolved with 24 grid cells per shell thickness, which corresponds to a shell thickness of 0.00152 pc within a spatial domain of (0.05 pc)$^3$. By contrast, in \citet{Pittard2013}, the blister growth is described with 2--3 grid cells per shell thickness, but over a larger spatial domain of (8 pc)$^2$, resulting in a shell thickness of approximately 0.13--0.19 pc.

\vspace{0.2cm}
For case $B_{12-10}$ we have performed a lower resolution convergence test, $B_{12-10-\textrm{Test}}$, listed in Table \ref{table1}. The results exhibit that the percentage of the surviving dust mass after 10 years post-explosion is  $\sim 76$\%. Therefore, agreements within 0.1\% for the energy and 0.3\% for the dust mass fraction were obtained.

\section{Modeling limitations} \label{Limitations}

Our calculations include the main processes influencing the evolution of dust grains in type II SN explosions, but we simplified some aspects due to complexity. For example, grid effects in our simulations due to the Cartesian geometry, as seen in the quadrilateral symmetry in Figs. \ref{CaseS200}, \ref{CaseB200}, \ref{CaseB12}, and \ref{CaseB12-15-20} (see also the 1D density profiles in Appendix \ref{AppendixD}), may influence the results, especially in later stages, as the grid’s geometry can impose artificial symmetries and distort the fluid dynamics. 

For instance, during the late-time SN shock-CSM interaction (final columns of Figures \ref{CaseS200}, \ref{CaseB200}, and \ref{CaseB12}), the CSM/SN composite remnant exhibits densities in the order of $10^3-10^{4.5}$, $10^{4.2}-10^{6.5}$, and $10^{6.5}-10^{8.9}$ cm$^{-3}$ for cases $S_{200}$, $B_{200}$, and $B_{12}$, respectively. The highest densities are primarily observed in the equatorial region, likely influenced by the Cartesian grid effects. We caution the readers that these high densities may not be appropriate for the conditions observed in evolved SN remnants at several years after the explosion.

The simulations were stopped when the eruption/SN composite remnants expanded to the boundaries of the computational domain (occurring at 250, 115 and 10 years for cases $S_{200}$, $B_{200}$, and $B_{12}$, respectively). Hence, we did not track dust destruction at later times. Nevertheless, this does not affect our conclusions as in all cases the dust mass fraction have reached constant, stable values. Indeed, we found nearly complete dust destruction in $S_{200}$ and $B_{200}$, and the stabilization of the dust mass fraction in all $B_{12}$ cases. 

\vspace{0.2cm}
Another point to address is that our calculations do not account for the optically thick regime at high densities ($\gtrsim 10^{10}$ cm$^{-3}$). While this approximation is necessary for accurately quantifying radiative cooling, it doesn't significantly impact the validity of our results as the regions of interest throughout the evolution of the eruption/SN composite remnant remain within this limit. Nevertheless, the obtained estimates of the dust mass fraction in this work should be treated as upper limits in all scenarios (especially in the $B_{12}$ cases).

\vspace{0.2cm}
Our model neglects for fluctuations in abundances once they are initially set relative to solar abundances. Nevertheless, behind the forward shock, where the majority of radiative cooling occurs, dust grains act as the primary coolants above gas temperatures $\gtrsim 10^6$ K \citep{OstrikerSilk1973,Dwekwerner1981,Dwek1987}. This implies that a considerable fraction of radiative cooling is primarily determined by the evolving dust-to-gas mass ratio and grain size distribution, rather than by specific abundances.

\vspace{0.2cm}
Finally, we assume collisional ionization equilibrium and electron-ion energy equipartition to account for radiative losses. 
These assumptions are justified by the high CSM densities of our simulated remnants, which significantly reduce the timescales for electron-ion energy equipartition ($t_{eq}$) and collisional ionization equilibrium ($t_{CIE}$). Lower CSM density values, however, are expected to introduce substantial deviations from equilibrium.
These assumptions may also break down in localized regions, particularly due to the rapid heating by shock waves, which can drive the system out of equilibrium. This is observed only in model $S_{200}$, where non-equilibrium conditions arise in zones immediately behind the shock wave (characterized by $t_{eq}/t_{SN} > 1$). Nonetheless, as discussed in Section \ref{sec_200}, the radiative cooling timescale ($t_{cool}$) remains long in these regions ($t_{eq}/t_{cool} \ll 1$).

\section{Conclusions} \label{conclusions}

By employing three-dimensional hydrodynamic simulations using the AMR code FLASH, we have developed the first attempt to understand the survival of dust grains formed after late-stage stellar eruptions, in relation to an eventual hydrogen--rich core-collapse SN explosion. This has been achieved by following the evolution of the progenitor's stellar wind, an erupted spherical or bipolar CSM, and the subsequent expansion of a SN remnant.

Various scenarios were examined based on the shape of the CSM created by the eruption, and the time elapsed prior to the SN explosion. Below, we summarize the main findings derived from our research:

\begin{itemize}
    \item Our results emphasize how CSM geometry affects the SN remnant evolution and the dust destruction efficiency. They also reveal the crucial role of eruption-SN delay time, unexpectedly favoring dust survival due to strong radiative cooling induced by high CSM densities
    when this delay time is shorter.
    \item Our models show that with a 200-year eruption-SN separation, only about 25\% of dust survives post-SN explosion in the spherical case up to the simulated time, while the CSM dust is almost entirely destroyed in the bipolar case. 
    \item Conversely, a 12-year eruption-SN separation in the bipolar case, allows a higher likelihood of dust survival as approximately half of the injected kinetic energy is radiated away, conserving between $\sim20$--75\% of the CSM dust mass fraction.
    \item The amount of energy radiated away upon the SNR-CSM interaction is $\sim(0.4$--$2.5)\times10^{51}$ erg. Consequently, as the SN forward shock turns radiative and weakens (which occurs to varying extents with both short and long eruption-SN gaps), we can anticipate that its eventual impact on the surrounding interstellar dust locked up within the encompassing wind-driven shell will be substantially reduced. 
    \item A weaker SN forward shock implies the development of a weaker SN reverse shock, thus we can also expect a reduced destruction of SN-condensed dust grains.
\end{itemize}

Consequently, this study sheds light on the survival of pre- and post-supernova dust grains, as well as interstellar dust grains, to the overall dust content of local and high-redshift galaxies.

\begin{acknowledgements}
The authors thank the anonymous referee for his/her helpful comments and suggestions which improved the quality of the paper. D.B.S-H. acknowledges support through a scholarship granted by SECIHTI-México. The authors thankfully acknowledge computer resources, technical advice and
support provided by Laboratorio Nacional de Supercómputo del Sureste de México
(LNS), a member of the SECIHTI national laboratories, with project No. 202501004C. SJ acknowledge the support provided by the Czech Academy of Sciences, through its Programme to Support Prospective Human Resources – Postdoctoral Fellows (PPLZ), contract number L100032351. SJ and RW acknowledge support by the institutional project RVO:67985815. {\it Software:} FLASH v4.6 \citep{Fryxelletal2000}, Numpy \citet{Numpy}, Wind \citep{Wunschetal2017}, \textsc{Cinder} \citep{MartinezGonzalezetal2018}, Matplotlib \citep{Matplotlib}, SciPy \citep{SciPy}.
\end{acknowledgements}

\bibliographystyle{aa}
\bibliography{master.bib}

\begin{appendix}

\section{Dynamical and cooling timescales of the stellar wind}\label{Appendix0}

To estimate the characteristic dynamical timescale, $t_{dyn}$, in the free wind region, we use the relation \citep{Silichetal2003}

\begin{equation}
t_{dyn}(R) = \int_{R_w}^R \frac{dx}{V_w} \approx \frac{R - R_w}{V_w},
\end{equation}

where $R_w$ is the wind injection radius, and $V_w$ is the nearly constant wind velocity outside the free wind zone.

The cooling time, $t_{cool}$, is determined by the balance between the radiated energy and the gas's thermal energy within a radius $R$ \citep{MacLowMcCray1988}. For a monoatomic gas with 1 He atom per each 10 H atoms, the mean mass per atom $\mu = 14/11 \, m_H$ and $f = 3/2$, the thermal energy is

\begin{equation}
E_\text{th}(R) = 4 \pi f k n_w R_w^3 \int_{1}^{R/R_w} T(x)\, dx,
\end{equation}

where $k$ is the Boltzmann constant, $n_w = \displaystyle \frac{\dot{M}_w}{4 \pi \mu R_w^2 V_w}$ is the gas number density at $r = R_w$, $x = R/R_w$, and $T(x)$ is the temperature at radius $R$.

The radiated energy beyond $R_w$ over time $t$ is

\begin{equation}
E_\text{rad}(R) = 4 \pi R_w^3 n_w^2 \int_{0}^{t_{cool}} dt \int_{1}^{R/R_w} \Lambda(T) \, x^{-2}\, dx,
\end{equation}

where $\Lambda(T)$ is the cooling function. Setting $E_\text{rad} = E_\text{th}$ gives $t_{cool}(R)$ as a function of $R$. Radiative cooling dominates when $t_{cool} < t_{dyn}$, that leads to the critical radius $R_\text{crit}$.

In the adiabatic wind solution \citep{ChevalierClegg1985}, the temperature follows a power-law profile

\begin{equation}
T(r) = T_w \left(\frac{r}{R_w}\right)^{-4/3}.
\end{equation}

The thermal energy in this case is thus

\begin{equation}
E_\text{th}(R) = 12 \pi f k n_w T_w R_w^{3} \left[1- \left(\frac{R}{R_w}\right)^{-1/3}\right].
\end{equation}

which simplifies for $R \gg R_w$ to

\begin{equation}
E_\text{th}(R) \approx 12 \pi f k n_w T_w R_w^{3}.
\end{equation}

Approximating the low-temperature cooling ($T_w \leq 10^4 \, \text{K}$, ionization fraction of 0.1) derived by \citet{DalgarnoandMcCray1972} by

\begin{align}
\Lambda(T) = 
\begin{cases} 
\Lambda_0^{(1)} T^{-5/2}, & 10 \leq T < 70 \, \text{K}, \\
\Lambda_0^{(2)} T^{-1/2}, & 70 \leq T \leq 10^4 \, \text{K},
\end{cases}
\end{align}

with parameters:

\begin{align}
    & \quad \Lambda_0^{(1)} = 3.137 \times 10^{-30}  \text{erg s}^{-1} \text{cm}^3 \text{K}^{5/2}, \\
    & \quad \Lambda_0^{(2)} = 1.437 \times 10^{-26}  \text{erg s}^{-1} \text{cm}^3  \text{K}^{1/2},
\end{align}

the condition $t_{cool} = t_{dyn}$ yields the critical radius and time:

\begin{equation}
R_\text{crit} \approx \left[\frac{4 \pi \mu f k T_w R_w^{2} V_w^2}{\dot{M}_w \Lambda(T_w)}\right],
\end{equation}

\begin{equation}\label{eq:tcrit}
t_\text{crit} \approx \left[\frac{4 \pi \mu f k T_w R_w^{2} V_w}{\dot{M}_w \Lambda(T_w)}\right].
\end{equation}

For the given values of $R_w$, $\dot{M}_w$, $V_w$ and $T_w$, radiative cooling is predicted to dominate at distances that lie beyond the computational domains considered in our simulations and the associated timescales are much longer than in those in our simulations.

\section{Initial conditions for the eruption ejecta} \label{AppendixA}
It is considered that the eruption ejecta density distribution can be described by

\begin{equation}
    \rho_{\textrm{e}}(r) = \frac{M_{\textrm{e}}}{R_{\textrm{e}}^3}f_{\textrm{e}}(r/R_{\textrm{e}}),
\end{equation}

where $r$ is the distance from the eruption site, $R_{\textrm{e}}$ is the eruption ejecta insertion radius, and $f_{\textrm{e}}(w_{\textrm{e}})=f_{\textrm{e}}(r/R_{\textrm{e}})$ is a structure function given by the piece-wise function defined in equation \eqref{erup_density}, with an inner power-law tenuous cavity up to a radius $R_{\textrm{sh,e}}$, and a thin,  constant density outer shell between $R_{\textrm{sh,e}}$ and $R_{\textrm{e}}$.

From continuity and mass conservation, $f_{\textrm{0,e}}$ in equation \eqref{erup_density} is:
\begin{equation} 
    f_{\textrm{0,e}} = f_k(1-w_{\textrm{sh,e}})^{-k}, \ \ \ f_{k} = \frac{1}{4\pi \lambda_{f_{\textrm{e}}}},
\end{equation}
with
\begin{eqnarray} \label{eq1}
   \lambda_{f_{\textrm{e}}} &=& \frac{16}{3} + \frac{2}{3}(1-w_{\textrm{sh,e}})^{-1.5} (12w_{\textrm{sh,e}}-3w_{\textrm{sh,e}}^2-8) \nonumber \\
   &+& \frac{1}{3}(1- w_{\textrm{sh,e}})^{-2.5}(1-w_{\textrm{sh,e}}^3),
\end{eqnarray} 
where we have considered $k$ equal to 2.5. The input parameters $E_{\textrm{k,e}}, v_{\textrm{max,e}}$ and $M_{\textrm{e}}$ are related by the normalization of the total energy via
\begin{equation}\label{Energy_eq_1}
    E_{\textrm{k,e}} = \frac{1}{2} \ M_{\textrm{e}}I_e \ v_{\textrm{max,e}}^2 \int_{0}^{1} w^4 f_{\textrm{e}}(w_{\textrm{e}}) dw,
\end{equation}
where
\begin{equation}
 I_e=2\pi\int_0^{\pi} F_{\varphi}^2 \sin (\varphi) d \varphi.
\end{equation}
By integrating Equation \ref{Energy_eq_1}, the maximum expansion velocity is expressed as
\begin{equation}
    v_{\textrm{max,e}} = \left[ \left( \frac{2E_{\textrm{k,e}}}{I_e M_{\textrm{e}}}\right)  \frac{1}{f_{k} \lambda_{v} + \frac{f_{0,\textrm{e}}}{5}(1-w_{\textrm{sh,e}}^{5})} \right]^{1/2},
\end{equation}
where
\begin{eqnarray}
   \lambda_{v} &=& \frac{256}{15} + \frac{2}{3} (1-w_{\textrm{sh,e}} )^{-1.5} - 8(1-w_{\textrm{sh,e}})^{-0.5} \nonumber \\ &-& 12(1-w_{\textrm{sh,e}})^{0.5}
    +\frac{8}{3}(1-w_{\textrm{sh,e}})^{1.5} 
    - \frac{2}{5}(1-w_{\textrm{sh,e}})^{2.5}.  
\end{eqnarray}

\section{Asymmetries in the stellar wind}
\label{AppendixB}
\begin{figure*}
    \centering
    \includegraphics[width=0.9\textwidth]{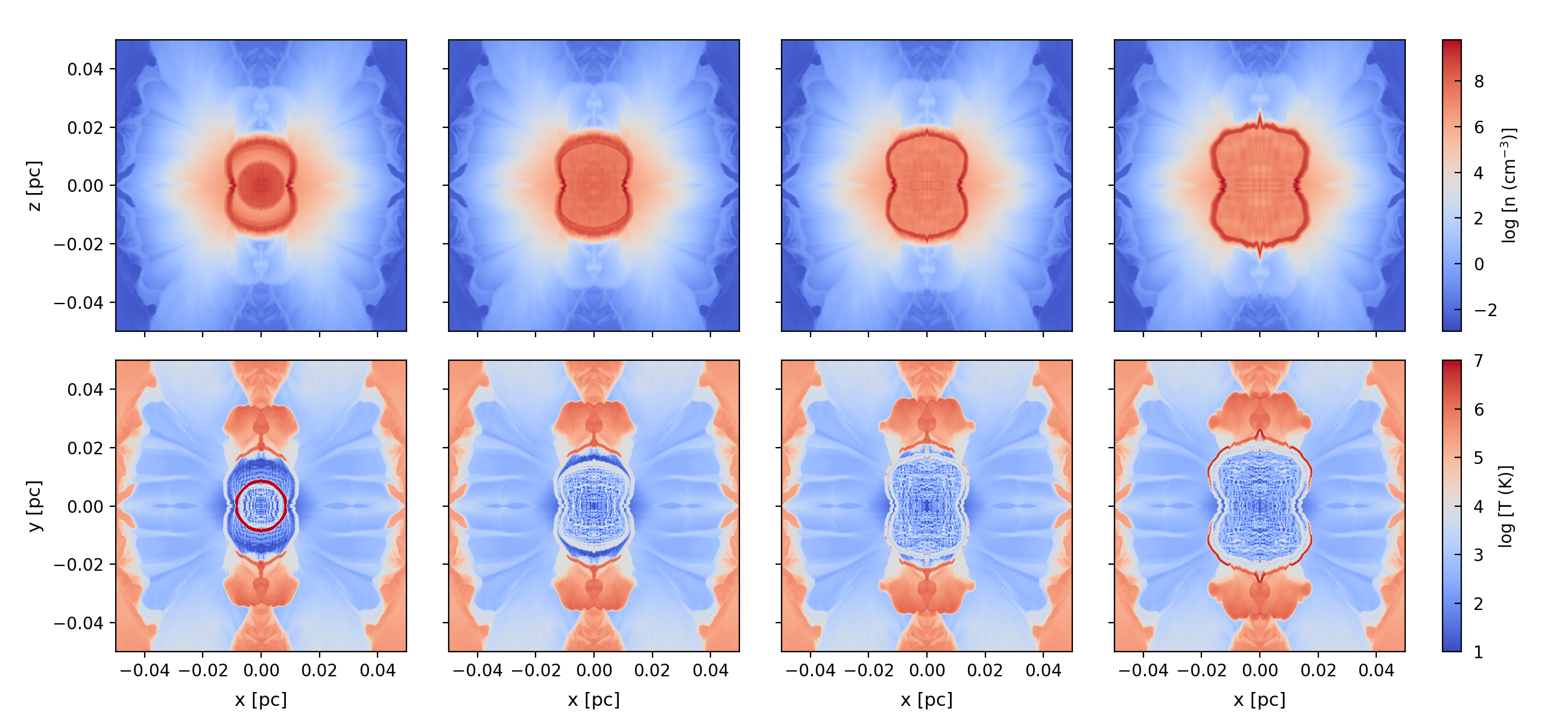}
    \captionsetup{skip=1pt}
    \caption{From left to right, the panels presents 1, 2, 5 and 7 yr after the supernova explosion. Note that the supernova shock becomes isothermal soon before colliding with the CSM shell thus allowing the survival of a large fraction of the dust introduced by the eruption. 
    }
    \label{CaseB12_wind}
\end{figure*}

It is also worth exploring the case in which the bipolar shape results from the interaction of a spherical eruption with a non-spherically symmetric stellar wind. Thus, we adopt the stellar wind model by \cite{Franketal1995}. Indeed, the density and velocity of the stellar wind are given by:
\begin{equation}
    \rho \left(r, \varphi \right)= \rho_0 \left( \frac{r_0}{r} \right)^2 \frac{1}{F \left(\varphi \right)} ,
\end{equation}
\begin{equation}
    v \left(r, \varphi \right)=v_{r,0} F \left(\varphi \right),
\end{equation}
where $F \left(\varphi \right)$ is the function defined in equation \eqref{Func_F}. 

Following \cite{Franketal1995}, model $B_{12-\textrm{wind}}$ considers a stellar wind consisting of a low-density component with $\alpha=0.995$, $\beta=2.3$, $\dot{M}_w=10^{-3}$ M$_\odot$ yr$^{-1}$, $v_{r,0}=250$ km s$^{-1}$, inserted within a radius of $r_0=0.006$ pc. The wind expands into a low density uniform ambient medium with density $\rho _{amb} \sim 10^{-27}$ g cm$^{-3}$. This component evolves during 290 years until filling most of the computational box of ($0.05$ pc)$^3$. Then, a second, denser, component is introduced for 20 years by increasing the mass input rate to $\dot{M}_w=10^{-1}$ M$_\odot$ yr$^{-1}$ and the velocity to $v_{r,0}=750$ km s$^{-1}$. Later, a spherically symmetric stellar eruption is injected as described in section \ref{sec_er} with $\alpha=0$. Finally, in order to compare this calculation with our model $B_{12-10}$ (see section \ref{sec_B12}), a supernova explodes 12 yr after the eruption with the same SN parameters as in $B_{12-10}$. 

Fig. \ref{CaseB12_wind} presents the simulation results, with panels from left to right showing snapshots at $t \sim 1, 2, 5$ and 7 yr after the supernova. First, note that a bipolar shape can indeed be the outcome of the interaction of a spherical eruption with the non-spherical wind. This occurs because the CSM formed by the stellar wind is significantly denser in the equatorial region, inhibiting the expansion of the eruption remnant in this direction and resulting in the bipolar morphology. The shape and size of the CSM shell prior to the SN are similar with those discussed in section \ref{sec_B12}, thus showing that our model setup (see section \ref{sec_er}) is able to describe the interaction of the eruption with the prior wind. This is even more relevant considering that here our aim is to study the outcome of the dust introduced by the eruption once the SN occurs. 

Fig. \ref{CaseB12_wind} shows that the SN forward shock reaches large temperatures at the earliest stages of evolution ($\sim 10^7$ K, see the first bottom panel in Fig. \ref{CaseB12_wind}). However, the shock soon becomes isothermal due to strong cooling and adopts the bipolar shape before colliding with the outer shell (middle panels in Fig. \ref{CaseB12_wind}), where most of the dust is located. Hence, similar to our $B_{12-10}$ case, most of the dust ($\sim55$\%, see Fig. \ref{dust-energy2}) survives even after the SN shock has overrun the CSM shell, which occurs at $t_{sn} \sim 7$ yr, as shown in the right columns of Fig. \ref{CaseB12_wind}.

\section{The role of radioactive heating}
\label{AppendixC}
The total energy released by the radioactive decay chain ${}^{56}\textrm{Ni}\rightarrow{}^{56}\textrm{Co}\rightarrow{}^{56}\textrm{Fe}$ can be approximated by \citep{Matsumotoetal2024}:
\begin{equation} \label{E56Ni}
E_{\textrm{Ni}} \simeq 1.8 \times 10^{48}  \left( \frac{M_{\textrm{Ni}}}{10^{-2} \, M_{\odot}} \right) \, \textrm{erg}  .
\end{equation}

For a typical value of $M_{\textrm{Ni}}\sim0.032$ \(M_{\odot}\) of \(^{56}\textrm{Ni}\) produced in type II SNe \citep{Anderson2019}, this yields approximately $5.76 \times 10^{48}$ erg. This energy, assumed to be evenly distributed and instantaneously transferred to heat a 10-M$_{\odot}$ ejecta (case $B_{12-\textrm{56Ni}}$), translates into a temperature $\sim2\times10^6$ K. For the assumed values in cases $B_{12-15}$ and $B_{12-20}$, the ejecta gas temperatures are $\sim4\times10^7$ K and $\sim8\times10^7$ K, respectively. 

\section{1D density profiles}
\label{AppendixD}
\begin{figure*}
    \centering
    \includegraphics[width=0.8\textwidth]{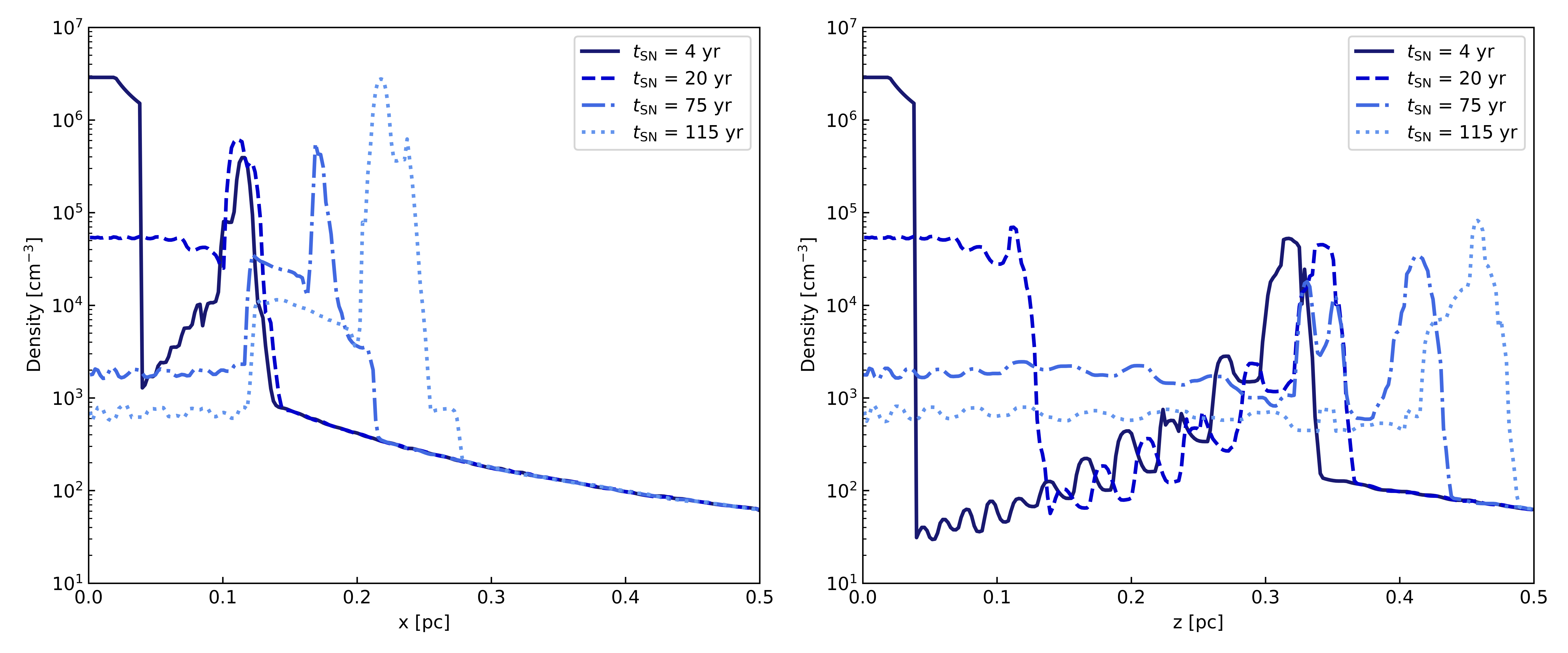}
    \captionsetup{skip=1pt}
    \caption{Density profiles along the x (left panel) and z (right panel) axes for $B_{200}$. The different curves correspond to the times presented in Fig. \ref{CaseB200}.}
    \label{DensityB200}
\end{figure*}

\begin{figure*}
    \centering
    \includegraphics[width=0.8\textwidth]{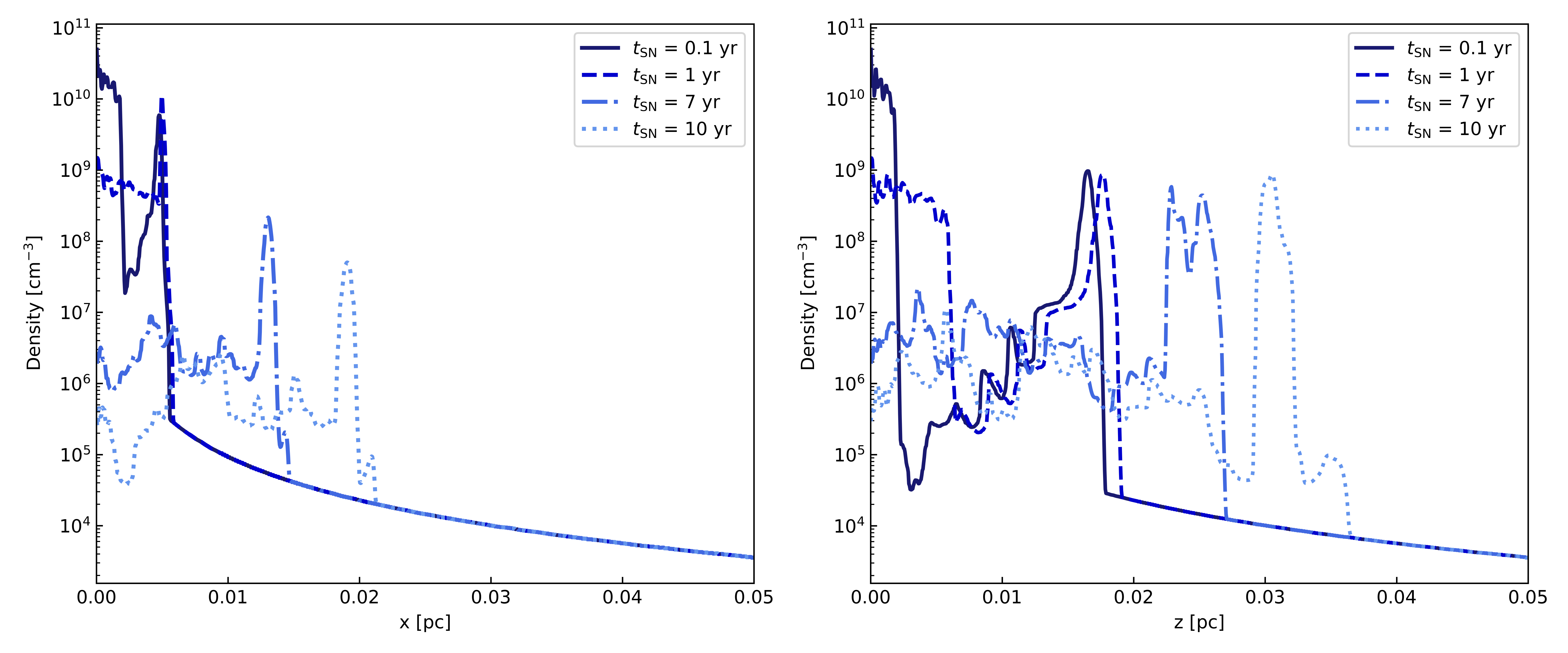}
    \captionsetup{skip=1pt}
    \caption{Same as Fig. \ref{DensityB200} but for $B_{12-10}$.}
    \label{DensityB12}
\end{figure*}

Figs. \ref{DensityB200} and \ref{DensityB12} show the density profiles in the x (left) and z (right) axes at the same post-SN times shown in Figs. \ref{CaseB200} and \ref{CaseB12} for the $B_{200}$ and $B_{12-10}$ cases, respectively. It is important to note that the stellar wind, the CSM shell created by the eruption, and the SN are all in agreement with the expected in our numerical setup (section \ref{section2}). Larger fluctuations in density are observed in the z-axis, which is due to the larger gas velocities in this direction.

\end{appendix}
\end{document}